\documentclass[fleqn,usenatbib]{mnras}

\usepackage{newtxtext} 

\usepackage[T1]{fontenc}

\DeclareRobustCommand{\VAN}[3]{#2}
\let\VANthebibliography\thebibliography
\def\thebibliography{\DeclareRobustCommand{\VAN}[3]{##3}\VANthebibliography}

\usepackage{graphicx}	
\usepackage{amsmath}	
\usepackage{amssymb}	

\usepackage{newtxmath}

\title[Molecular gas cloud properties at $z\simeq 1$]{Molecular gas cloud properties at $z\simeq 1$ revealed by the superb angular resolution achieved with ALMA and gravitational lensing}

\author[M. Dessauges-Zavadsky et al.]{
Miroslava Dessauges-Zavadsky,$^{1}$\thanks{E-mail: miroslava.dessauges@unige.ch}
Johan Richard,$^{2}$
Fran\c{c}oise Combes,$^{3}$
Matteo Messa,$^{1,4}$ 
\newauthor 
\hspace{1pt} David Nagy,$^{1}$
Lucio Mayer,$^{5}$
Daniel Schaerer,$^{1,6}$
Eiichi Egami$^{7}$
and Angela Adamo$^{8}$
\\
$^{1}$D\'epartement d'Astronomie, Universit\'e de Gen\`eve, Chemin Pegasi 51, 1290 Versoix, Switzerland\\
$^{2}$Universit\'e Lyon, Universit\'e Lyon1, Ens de Lyon, CNRS, Centre de Recherche Astrophysique de Lyon UMR5574, 69230 Saint-Genis-Laval, France\\
$^{3}$Observatoire de Paris, LERMA, Coll\`ege de France, CNRS, PSL University, Sorbonne University, 75014 Paris, France\\
$^{4}$The Oskar Klein Centre, Department of Astronomy, Stockholm University, AlbaNova, 10691 Stockholm, Sweden\\
$^{5}$Center for Theoretical Astrophysics and Cosmology, Institute for Computational Science, University of Zurich, Winterthurerstrasse 190, 8057 Z\"urich, Switzerland\\
$^{6}$CNRS, IRAP, Avenue E. Belin 14, 31400 Toulouse, France\\
$^{7}$Steward Observatory, University of Arizona, 933 N. Cherry Avenue, Tucson, AZ 85721, USA\\
$^{8}$The Oskar Klein Centre, Department of Astronomy, Stockholm University, AlbaNova, 10691 Stockholm, Sweden
}

\date{Accepted XXX. Received YYY; in original form ZZZ}

\pubyear{2015}

\begin{document}
\label{firstpage}
\pagerange{\pageref{firstpage}--\pageref{lastpage}}
\maketitle

\begin{abstract}
Current observations favour that the massive ultraviolet-bright clumps with a median stellar mass of $\sim 10^7~M_{\odot}$, ubiquitously observed in $z\sim 1-3$ galaxies, are star-forming regions formed in-situ in galaxies. It has been proposed that they result from gas fragmentation due to gravitational instability of gas-rich, turbulent, high-redshift discs. We bring support to this scenario by reporting the new discovery of giant molecular clouds (GMCs) in the strongly lensed, clumpy, main-sequence galaxy, A521-sys1, at $z=1.043$. Its CO(4--3) emission was mapped with the Atacama Large Millimeter/submillimeter Array (ALMA) at an angular resolution of $0.19''\times 0.16''$, reading down to 30~pc thanks to gravitational lensing. We identified 14 GMCs, most being virialized, with $10^{5.9}- 10^{7.9}~M_{\odot}$ masses and a median $800~M_{\odot}~\mathrm{pc}^{-2}$ molecular gas mass surface density, that are, respectively, 100 and 10 times higher than for nearby GMCs. They are also characterized by 10 times higher supersonic turbulence with a median Mach number of 60. They end up to fall above the Larson scaling relations, similarly to the GMCs in another clumpy $z\simeq 1$ galaxy, the Cosmic Snake, although differences between the two sets of high-redshift GMCs exist. Altogether they support that GMCs form with properties that adjust to the ambient interstellar medium conditions prevalent in the host galaxy whatever its redshift. The detected A521-sys1 GMCs are massive enough to be the parent gas clouds of stellar clumps, with a relatively high star-formation efficiency per free-fall time of $\sim 11$ per cent.
\end{abstract}




\section{Introduction}

The understanding of the process of star formation at high redshift, which drives the stellar mass build-up of galaxies across cosmic time, evolves in synergy with better observational facilities and not only the improvement of their sensitivity, but specifically their spatial resolution. The {\it Hubble Space Telescope} ({\it HST}) has delivered unprecedented, resolved near-ultraviolet (UV) to near-infrared (IR) images of high-redshift galaxies. They revealed that about 60 per cent of star-forming galaxies around the peak of the cosmic star-formation rate density ($1<z<3$) have perturbed morphologies, dominated by giant UV-bright star-forming knots/clumps \citep[e.g.][]{Cowie95,Conselice04,Elmegreen07,Elmegreen13,Livermore12,
Livermore15,Guo12,Guo15,Guo18,Soto17,Zanella19}. The origin and nature of these clumps have been debated over two decades now, while their understanding could provide key information on:
the mass assembly of galaxies, via mergers or secular evolution, depending on whether the clumps we see are 'ex-situ' satellites resulting from a merger event or are formed locally ('in-situ') in galaxies; the structural evolution of the host galaxy if clumps contribute to the bulge growth when migrating toward the galactic center; the star-formation process of high-redshift galaxies if clumps are star-forming regions formed in-situ in galaxies; and the role clumps play in the galactic stellar feedback and Lyman continuum ionizing photons \citep{Dessauges20}.

Several observational findings currently support that the observed clumps are star cluster complexes/large star-forming regions formed in-situ in high-redshift galaxies. We can cite the study by \citet{Shibuya16} of 17\,000 galaxies from the {\it HST} legacy surveys, which evidenced that the redshift evolution of the clumpy galaxy fraction differs from the one of both minor and major mergers \citep{Lotz11}. The same authors also showed that the bulk of the clumpy galaxies are disc-like systems with a S\'ersic index close to 1. This is consistent with their kinematics found to be dominated by ordered disc rotation with high velocity dispersions, similarly to the majority of star-forming galaxies at $0.8<z<3$ \citep[e.g.][]{Forster09,Wisnioski15,Turner17,Simons17,Ubler19,Girard18,Girard20}. Moreover, \citet{Elmegreen17} determined that the scale-heights of clumpy galaxies and those of normal spirals when seen edge-on are comparable, indicating that clumps are distributed within the disc thickness of the host galaxy \citep[see also][]{ElmegreenEl06}. Finally, \citet{Dessauges18} reported that the stellar mass function of clumps at $1<z<3$ follows a power-law with a slope close to $-2$, similar to the one of local/nearby star clusters and H\,{\sc ii} regions \citep[e.g.][]{Grasha17,Adamo17,Messa18}, and expected if clumps are formed via the turbulence-driven fragmentation process \citep{Elmegreen06,Guszejnov18,Ma20}.

The compilation of UV-bright clumps hosted the {\it HST} Ultra Deep Field galaxies and in the strongly lensed galaxies, now also imaged with the {\it James Webb Space Telescope} ({\it JWST}), which benefits from the best sensitivity and the best spatial resolution imaging, shows clump sizes $\lesssim 300~\mathrm{pc}$ and stellar masses ranging from $\sim 10^5$ to $\sim 10^9~\mathrm{M}_{\odot}$ with a median of $\sim 10^7~\mathrm{M_{\odot}}$ \citep{Elmegreen13,Dessauges17,Cava18,Mestric22,Messa22,Claeyssens22}. These revised clump properties, indicating sizes much smaller than 1~kpc and masses with a median lower than $10^{8.5-9}~\mathrm{M}_{\odot}$ initially reported, nevertheless, confirm that the high-redshift clumps remain more massive by $2-3$ orders of magnitude than the local/nearby star clusters \citep[e.g.][]{Bastian06,Adamo13}. They yet have stellar mass surface densities comparable to the average surface density of a nearby massive cluster \citep{Brown21,Messa22,Claeyssens22}. They are very well reproduced by the high-resolution hydrodynamical simulations \citep[e.g.][]{Tamburello15,Mandelker17,Fensch21}, which have proposed that the high-redshift, gas-rich, rotation-dominated galaxies, characterized by highly turbulent gas discs, fragment because of violent gravitational instability driven by intense cosmic web cold gas inflows \citep[e.g.][]{Agertz09,Dekel09,Bournaud10}, and produce gravitationally bound gas clouds, the progenitors of the stellar clumps ubiquitously seen in high-redshift galaxies. If this scenario is correct, one should also observe the parent gas clouds.

\begin{figure*}
\includegraphics[width=15.5cm,clip]{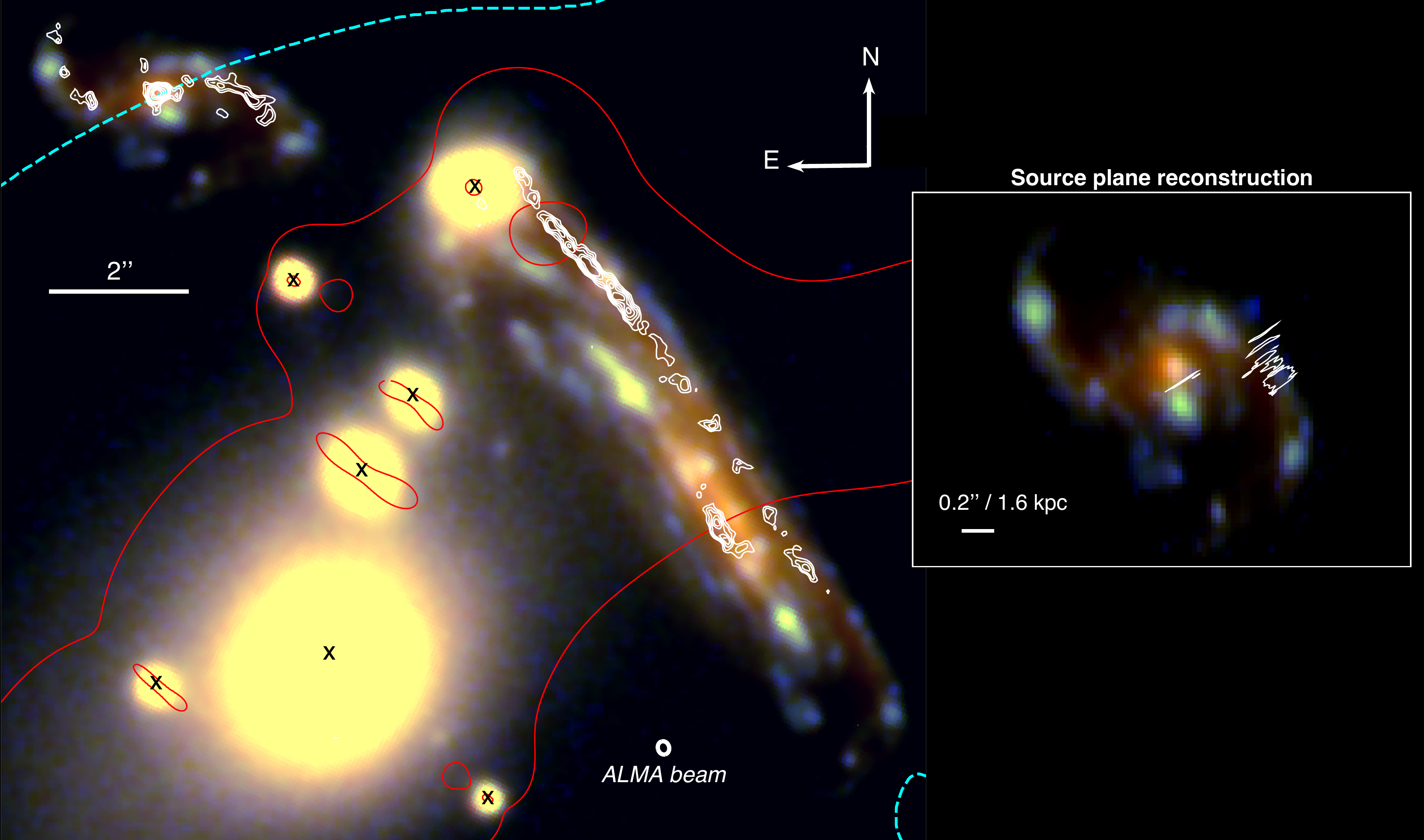}
\caption{{\it HST} RGB-colour composite image (red filter: F160W; green: F105W; blue: F606W) of the multiply-imaged A521-sys1 galaxy together with the foreground Abell~521 galaxy cluster. The black crosses mark the cluster members. The critical line of our lens model at $z=1.043$, the redshift of the A521-sys1 galaxy, is shown by the red solid line. The entire A521-sys1 galaxy can be seen in the north-east isolated and uniformly lensed image, which reveals the spiral structure of the galaxy with two spiral arms. The 11~arcsec-long arc in the western direction is a five-fold multiple image of the A521-sys1 galaxy, with two main mirrored images encompassing the critical line, and the remaining multiple images covering a small region around the closed critical line in the north of the arc. Only about 55 per cent of the galaxy, including less than half of the galactic central region and one spiral arm (the western arm of the isolated counter-image), is multiply-imaged along the arc, as shown by the cyan dashed line that delimits southward the zone of multiple images. The overlaid white contours correspond to the {\it ALMA} CO(4--3) velocity-integrated intensity in levels of $4\sigma$, $5\sigma$, $6\sigma$, $7\sigma$, $8\sigma$, $10\sigma$, $12\sigma$, and $14\sigma$, with the RMS noise of $0.01~\mathrm{Jy~beam^{-1}~km~s^{-1}}$. The {\it ALMA} synthesized beam with a size of $0.19''\times 0.16''$ at the position angle of $-74^{\circ}$ is represented by the white ellipse. We give a reference scale of 2~arcsec. The left inset shows the {\it HST} RGB-colour composite source-plane reconstruction of the lensed isolated image with the source-plane reconstructed CO(4--3) velocity-integrated intensity $4\sigma$ contours (in white) detected in the arc. The 14 identified GMCs are not resolved in the source-plane image, they are distributed within the CO(4--3) $4\sigma$ contours.}
\label{fig:HST-ALMA}
\end{figure*}

The cold molecular gas in nearby galaxies is usually traced by the carbon monoxide (CO) and is structured in giant molecular clouds (GMCs). These GMCs with $10^4-10^7~\mathrm{M}_{\odot}$ masses and $5-100$~pc radii contract and form star clusters \citep{Bolatto13}. Their detection at such a small scale in distant galaxies is observationally challenging even in interferometric data. Numerous studies searching for these clouds in the far-IR dust continuum images of high-redshift submillimeter galaxies (SMGs) led to non-detections showing essentially a smooth dust emission \citep[e.g.][]{Hodge16,Hodge19,Gullberg18,Rujopakarn19,Ivison20}. 
Only a few studies reported GMC detections in the CO emission maps of SMGs \citep{Swinbank15,Canameras17,Tadaki18}. We obtained the first detection of GMCs in a clumpy, main-sequence, rotation-dominated galaxy---the strongly lensed Cosmic Snake galaxy at $z=1.036$---characterized by a stellar mass of $M_{\mathrm{stars}} = (4.0\pm 0.5)\times 10^{10}~\mathrm{M}_{\odot}$, a star-formation rate of $\mathrm{SFR} = 30\pm 10~\mathrm{M_{\odot}~yr^{-1}}$, and a molecular gas to stellar mass fraction of $f_{\mathrm{molgas}} = 25\pm 4$ per cent \citep{Dessauges19}. The kinematics, the metallicity, and the radial profiles of surface densities of $M_{\mathrm{stars}}$, $\mathrm{SFR}$, molecular gas mass ($M_{\mathrm{molgas}}$), and dust mass of the Cosmic Snake galaxy were studied by \citet{Patricio18,Patricio19}, \citet{Girard19}, and \citet{Nagy22}, respectively. The combination of the superb angular resolution achieved with {\it ALMA} and gravitational lensing allowed to map the emission of the CO(4--3) line of this galaxy at the resolution of nearby galaxies \citep{Sun18,Rosolowsky21}, reading down to 30~pc, and to perform the GMC search in the velocity space (in channel intensity maps). We identified 17 GMCs 
with on average 100 times higher $M_{\mathrm{molgas}}$ and 10 times higher $M_{\mathrm{molgas}}$ surface densities than for most of the local/nearby GMCs \citep{Dessauges19}. The high masses of the identified GMCs are close to the stellar masses of the UV-bright clumps detected by \citet{Cava18} in comparable numbers in the {\it HST} images of the Cosmic Snake galaxy at a similar spatial resolution to the {\it ALMA} images. In \citet{Dessauges19} we concluded that this corroborates the in-situ formation of high-redshift stellar clumps. 
Moreover, we proposed that the physical properties derived for the Cosmic Snake GMCs (high mass, high density, and high internal turbulence) suggest that, also at high redshift, GMCs seem to form with properties that adjust to the host galaxy's ambient interstellar medium (ISM) conditions (pressure, turbulence, density, etc.), as observed for nearby GMCs \citep{Hughes13,Sun18,Sun20}.

In this work we study another clumpy, main-sequence, proto-typical Milky Way galaxy---the A521-sys1 galaxy at $z=1.043$---strongly lensed by the galaxy cluster Abell~521. Its kinematics is characteristic of rotation-dominated galaxies at cosmic noon, with a high velocity dispersion and a low rotation velocity over velocity dispersion ratio of $\sim 2$ \citep{Patricio18,Girard19}. Resembling 
the Cosmic Snake galaxy, in terms of redshift, $M_{\mathrm{stars}} = (7.4\pm 1.2)\times 10^{10}~\mathrm{M}_{\odot}$, $\mathrm{SFR} = 26\pm 5~\mathrm{M_{\odot}~yr^{-1}}$ \citep{Nagy22}, and $f_{\mathrm{molgas}} = 14\pm 3$ per cent (this work), we seek to test and generalize the results found for the Cosmic Snake GMCs. A detailed analysis of the {\it HST} near-UV to near-IR images by \citet{Messa22} allowed to identify 18 UV-bright clumps with stellar masses of $1.5\times 10^6-6\times 10^8~\mathrm{M}_{\odot}$. Here we analyse the {\it ALMA} CO(4--3) observations obtained at a spatial resolution close to the {\it HST} images, which we use to search for GMCs down to a physical scale of $\sim 30$~pc. 

The paper is organized as follows. In Section~\ref{sect:lens-model} the lens model of Abell~521 is presented. Section~\ref{sect:observations} describes the {\it ALMA} observations, and the data reduction and imaging. In Section~\ref{sect:analysis} we explain how the search for the molecular clouds was performed and how their physical properties were derived. We discuss the GMC physical properties and their implications in Section~\ref{sect:discussion}. Our conclusions can be found in Section~\ref{sect:conclusions}. We adopt the $\Lambda$-CDM cosmology with $H_0 = 70~\mathrm{km~s^{-1}~Mpc^{-1}}$, $\Omega_{\mathrm{M}} = 0.3$, $\Omega_{\Lambda} = 0.7$, and the \citet{Salpeter55} initial mass function.


\section{Lens model}
\label{sect:lens-model} 

Lensed by the galaxy cluster Abell~521, the targeted galaxy A521-sys1 is multiply-imaged as illustrated in Fig.~\ref{fig:HST-ALMA}. There is one isolated and uniformly lensed image of the entire galaxy, magnified by a factor $\mu = 2$ to 6, observed in the north-east; and five strongly magnified and stretched images, covering about 55 per cent of the galaxy, that form a wide arc in the western direction of 11~arcsec in length from north to south, with high $\mu$ ranging from 3 to $>50$ and even $>100$ in sub-regions close to the critical line \citep[see][Fig.~2 for a magnification map]{Nagy22}. The former image reveals the spiral structure of the galaxy with two spiral arms, 
while the latter images provide a zoom-in on the internal structure of less than half of the galaxy 
including the western spiral arm of the former image. We use the tailored lens model of the A521 cluster from \citet{Richard10}, which was refined to better match the arc at $z=1.043$ by using 7 multiply-imaged UV-bright clumps as constraints \citep[see][for details]{Messa22}. The resulting root-mean-square (RMS) noise between the observed and predicted locations measured in the image-plane is 0.08~arcsec, and $\mu$ is determined with a typical uncertainty of the order of $\sim 10-20$ per cent. The reconstructed source-plane image of the A521-sys1 galaxy is shown in the left inset of Fig.~\ref{fig:HST-ALMA}.


\section{ALMA observations, data reduction, and imaging}
\label{sect:observations}

The A521-sys1 galaxy was observed in Cycle~4 (2016.1.00643.S) in band~6 with 41 {\it ALMA} 12~m antennae in the C40-6 configuration (maximum baseline of 3.1~km) during 89~min of on-source time. The redshifted CO(4--3) line frequency at 225.6685~GHz was used to tune the spectral window 2, which spectral resolution was set to 7.8125~MHz ($10.3782~\rm km~s^{-1}$). We used the remaining three 1.875~GHz spectral windows for continuum emission. As described in \citet{Girard19}, the {\it ALMA} data were reduced following the reduction pipeline distributed within CASA \citep[the Common Astronomy Software Application;][]{McMullin07}. For the CO(4--3) line imaging, we used a pixel size of $0.03$ arcsec and the Briggs weighting with the robust factor of 0.5, and we cleaned all channels until convergence using the \texttt{clean} routine in CASA with a custom mask. According to the {\it ALMA} Technical Handbook, the adopted weighting scheme is known to deliver low sidelobe levels for $\sim 1$~h observations and to give a reasonable balance between spatial resolution and sensitivity. We finally applied the primary beam correction. The synthesized beam size of the resulting CO(4--3) line data cube is $0.19''\times 0.16''$ at the position angle of $-74^{\circ}$. The cube has an RMS noise of $0.2~\rm mJy~beam^{-1}$ per $10.3782~\rm km~s^{-1}$ channel. The calibrated visibilities of continuum over the four spectral windows, when excluding channels where the CO(4--3) emission was detected, were also imaged, but no 1.3~mm continuum was detected. 

The \texttt{immoments} routine in CASA was used to obtain moment maps of the CO(4--3) emission. The threshold was fixed to $4\sigma$ RMS when computing the velocity first-moment and the velocity dispersion second-moment. We used the same routine to extract 12-channel intensity maps (zero-moment maps), at the native spectral resolution of $10.3782~\rm km~s^{-1}$, of the CO(4--3) emission covering the velocity range from $-41$ to $+73~\rm km~s^{-1}$; the zero velocity was set to $z = 1.043425$. The respective channel maps are plotted in Fig.~\ref{fig:Appendix}. 

In Fig.~\ref{fig:HST-ALMA} we overlay the contours of the CO(4--3) velocity-integrated intensity on the {\it HST} RGB-colour composite image of the A521-sys1 galaxy. The two spiral arms nicely visible in the {\it HST} near-UV to near-IR images of the north-east isolated and uniformly lensed image are also detected in the CO(4--3) emission, together with the CO-bright galactic central region. The western spiral arm is clearly more gas-rich than the eastern spiral arm. By chance, the western spiral arm is the arm that is multiply-imaged over the arc, where it benefits from high magnification factors.
\citet{Nagy22} showed that the CO(4--3) emission extends out to the galactocentric radius of $\sim 6$~kpc in the A521-sys1 source galaxy. This corresponds to $\sim 1/3$ of the radial extent of the stellar emission ($\sim 20$~kpc) in the rest-frame optical {\it HST} F160W filter. Asymmetries in the molecular gas distribution between two spiral arms and in scale lengths of the stellar and molecular gas emission radial profiles are common in galaxies: they can be inherited from the cosmic web accretion pattern, but they can also be due to internal instabilities and the fact that dark matter halos are not spherical and can produce warps, bending modes, etc. Finally, asymmetries can be induced by interactions with (big) satellites. In the case of the A521-sys1 galaxy, there is no evidence of a massive satellite at small projected distance.


\section{Analysis}
\label{sect:analysis}

\subsection{Search for molecular clouds}
\label{sect:GMC-search}

To search for GMCs in the A521-sys1 galaxy, we considered only the multiple images within the arc, because there the magnification factors are the higher (yielding to a better spatial resolution) and because the arc is located close to the phase center, such that the RMS noise over the arc's CO(4--3) emission is almost constant, affected by less than 4 per cent by the primary beam attenuation.
We followed the methodology developed in \citet{Dessauges19} for the Cosmic Snake galaxy, which consisted in exploiting the three dimensions of the CO(4--3) data cube by searching for emission peaks of the molecular clouds in the 12-channel intensity maps extracted over the CO(4--3) emission (Fig.~\ref{fig:Appendix}). The reliability of an emission detection against noise was evaluated by computing the fidelity of the detection at a given significance:
\begin{equation}
\mathrm{fidelity(S/N)} = 1-\frac{N_{\mathrm{neg}}(\mathrm{S/N})}{N_{\mathrm{pos}}(\mathrm{S/N})},
\label{eq:fidelity}
\end{equation}
where $N_{\mathrm{pos}}$ and $N_{\mathrm{neg}}$ are, respectively, the number of positive and negative emission detections with a given signal-to-noise ratio (S/N) in the primary beam \citep{Walter16,Decarli19}. In individual channel maps the fidelity of 100 per cent was reached at $\mathrm{S/N} = 4.8$, thus we first extracted, in the primary beam, all $>4.8\sigma$ emissions per channel. Then, we considered emissions spatially overlapping in two adjacent $10.3782~\rm km~s^{-1}$ channel maps, for which the fidelity of 100 per cent was reached at $\mathrm{S/N} = 3.9$ per channel (equivalent to $\mathrm{S/N} = 5.5$ for two adjacent channels). Therefore, we also extracted all $>3.9\sigma$ emissions co-spatial in at least two adjacent channel maps\footnote{We considered two CO(4--3) emissions as co-spatial in two adjacent channel maps when their respective $4\sigma$ contours were overlapping over at least three pixels of 0.03~arcsec.}. All the extracted CO(4--3) emissions were found to be distributed over the position of the detected CO(4--3) velocity-integrated intensity of the A521-sys1 multiple images, which contours are shown in white in Fig.~\ref{fig:HST-ALMA}. This provided support that the extracted emissions were detected at sufficiently high S/N to be real, and were not noise peaks since they were not randomly distributed. As an additional test, we also performed the search for cloud's emission peaks in the `dirty' channel maps (before the \texttt{clean}) that gave the same results as the emission search in the cleaned channel maps. This allowed us to establish that the \texttt{clean} did not give rise to noise peaks, or artificial cloud detections.

Similarly to the analysis performed for the Cosmic Snake galaxy \citep{Dessauges19}, we adopted the same definition of molecular clouds by considering CO(4--3) emissions as belonging to distinct clouds when their $4\sigma$ contours were not spatially overlapping in the same channel map, or when their $4\sigma$ contours were co-spatial, but not in adjacent channel maps (i.e.\ when separated by at least one channel). This allows clouds to overlap in either velocity or physical space, but not both. Following the above definition, with the help of the lens model, we could group the extracted CO(4--3) emissions spread over the A521 arc in 24 multiple images of 14 distinct GMCs. 
We found no GMCs overlapping along the same line of sight,
and almost all GMCs (except the GMCs with IDs 1 and 2) are multiply-imaged (twice to three times) in the 12-channel intensity maps analysed as shown in Fig.~\ref{fig:Appendix}. In fact, all the GMCs identified in one single channel map (the GMCs with IDs 7, 9, and 12), except the GMC~6, are detected in the two multiple images predicted by the lens model in the arc. The detection of these respective multiple images at the predicted locations reinforces the reliability of these GMCs, since it is unlikely to find two noise features sitting at these predicted locations. In the case of the GMC~6, we detect only one multiple image that, in reality, is a blend (within the synthesized beam) of two predicted multiple images because located almost on the critical line (see Fig.~\ref{fig:Appendix}; same for the GMC~13).
The detection of multiple images also enables us to alleviate the caveat of the possible effect of the Gibbs phenomenon, which is known to produce a non-trivial correlation across two adjacent channels resulting from the post-processing smoothing (e.g. Hanning; see the {\it ALMA} Technical Handbook). Again, all the GMCs identified over less than 3 adjacent channel maps that could be affected by the Gibbs phenomenon (the GMCs with IDs 10 and 14), are detected in two multiple images. Finally, in Appendix~B we discuss the results of the CASA visibility simulations we performed that exclude the possibility that the identified GMCs are artificial clumpy features produced by lensing and interferometric effects out of a smooth exponential disc of CO(4--3) emission.

The significance level of the 14 identified GMCs ranges between $\sim 5\sigma$ and $\sim 14\sigma$ (Table~\ref{tab:GMCs}) when integrated over their extent in velocity space from one to five adjacent $10.3782~\rm km~s^{-1}$ channels. The GMCs with IDs 10 to 14 are found near the galactic center, and the GMCs with IDs 1 to 10 are spread along the spiral arm (the western arm of the uniformly lensed image; see Fig.~\ref{fig:HST-ALMA}). 

The 14 GMCs are spatially offset from the 13 UV-bright stellar clumps detected over the arc by \citet[][see their Fig.~2]{Messa22}. The closest GMC/stellar clump pairs, located both near the galactic center and in the spiral arm, have a separation which ranges between $\sim 110$ and $\sim 980$~pc, with a median/mean of $\sim 590$~pc. This represents quite a large drift for the young stars from their birth place given the estimated ages of the stellar clumps < 100~Myr (the majority even have best-fit ages close to $10$~Myr, although with large uncertainties), much shorter than the dynamical time of the host galaxy \citep{Messa22}. The association between the identified GMCs and stellar clumps can thus be questionable. 
Considering the typical lifetime of GMCs in local galaxies \citep[$\sim 10-30$~Myr;][]{Kruijssen19,Chevance20,Kim21}, we can also wonder if the stellar clumps we are observing in A521-sys1 are not already old enough to have destroyed/dispersed their parent GMCs after the episode of star formation, unless high-redshift GMCs are more long-lived than the local ones given their different physical properties (see Sect.~\ref{sect:discussion}).

\begin{table}
\centering
\caption{Physical properties of the 14 GMCs identified in the A521-sys1 galaxy. The description of their measurements can be found in Sect~\ref{sect:GMC-properties}. $M_{\mathrm{molgas}}$ is the GMC lensing-corrected molecular gas mass obtained by adopting the CO-to-H$_2$ conversion factor, $\alpha_{\mathrm{CO}} = 1.9\pm 0.4~\mathrm{M_{\odot}(K~km~s^{-1}~pc^2)^{-1}}$, derived in Sect.~\ref{sect:virial}, $R$ is the GMC lensing-corrected beam-deconvolved circularised FWHM radius, $\sigma_{v}$ is the GMC interval velocity dispersion, and $\mu$ is the GMC mean magnification factor. The listed values were measured on the multiple images with the less blended CO(4--3) emissions in channel intensity maps (see Fig.~\ref{fig:Appendix}). The GMCs with IDs 1 to 10 are spread along the spiral arm, and the GMCs with IDs 11 to 14 are located near the galactic center. The letters are used to refer to the different multiple images of a given GMC, labelled following the nomenclature explained in Appendix~A and shown in Fig.~\ref{fig:Appendix}.}
\label{tab:GMCs}
\begin{tabular}{l r@{ $\pm$ }l r@{ $\pm$ }l r@{ $\pm$ }l r@{ $\pm$ }l r@{.}l} 
\hline
GMC & \multicolumn{2}{c}{$M_{\mathrm{molgas}}$} & \multicolumn{2}{c}{$R$} & \multicolumn{2}{c}{$\sigma_{v}$} & \multicolumn{2}{c}{$\mu$} & \multicolumn{2}{l}{Detection} \\
ID & \multicolumn{2}{c}{$\mathrm{10^6~M_{\odot}}$} & \multicolumn{2}{c}{pc} & \multicolumn{2}{c}{$\mathrm{km~s^{-1}}$} & \multicolumn{2}{c}{} & \multicolumn{2}{l}{significance} \\
\hline
1N  & $77$&$11$  & $166$&$16$ & $21$&$4$ & $18$&$4$   & $10$&$7\sigma$ \\ 
2N  & $66$&$12$  & $129$&$21$ & $17$&$4$ & $24$&$3$   & $12$&$0\sigma$ \\ 
3CS & $44$&$7$   & $144$&$37$ & $17$&$3$ & $26$&$4$   & $9$&$7\sigma$  \\ 
4CS & $18$&$4$   & $89$&$42$  & $11$&$4$ & $66$&$14$  & $8$&$5\sigma$  \\ 
5N  & $37$&$5$   & $104$&$38$ & $25$&$8$ & $30$&$3$   & $9$&$0\sigma$  \\ 
6C  & $0.8$&$0.1$& $31$&$6$   & $20$&$6$ & $366$&$38$ & $4$&$8\sigma$  \\ 
7S  & $28$&$5$   & $107$&$15$ & $16$&$4$ & $9$&$2$    & $6$&$8\sigma$  \\ 
8N  & $61$&$10$  & $117$&$19$ & $12$&$5$ & $16$&$3$   & $9$&$4\sigma$  \\ 
9S  & $6$&$1$    & $94$&$14$  & $8$&$2$  & $48$&$7$   & $7$&$2\sigma$  \\ 
10S & $32$&$5$   & $112$&$19$ & $5$&$2$  & $20$&$4$   & $7$&$5\sigma$  \\ 
11S & $41$&$5$   & $94$&$25$  & $16$&$4$ & $78$&$24$  & $13$&$9\sigma$ \\ 
12S & $15$&$2$   & $102$&$23$ & $11$&$3$ & $33$&$5$   & $6$&$9\sigma$  \\ 
13NS& $6$&$1$    & $28$&$7$   & $15$&$4$ & $428$&$147$& $12$&$4\sigma$ \\ 
14S & $22$&$3$   & $140$&$29$ & $14$&$4$ & $35$&$5$   & $6$&$7\sigma$  \\ 
\hline
\end{tabular}
\end{table}

\subsection{Physical properties of molecular clouds}
\label{sect:GMC-properties}
 
The molecular gas mass of each multiple image of a given GMC was derived by summing the lensing-corrected line-integrated fluxes of all the CO(4--3) emissions extracted for that image over a number of adjacent channels. To measure the fluxes, we used custom apertures per channel, large enough to include for each emission, all the signal above the local RMS noise level (no aperture correction was needed as the apertures used were all larger than the synthesized beam). To correct for the lensing, we considered the mean magnification factor computed over the area subtended by the $4\sigma$ intensity contour of each emission per channel\footnote{Given the small sizes of the $4\sigma$ intensity contours of the CO(4--3) emissions per channel, barely more extended in surface than the synthesized beam (at most by a factor of 1.7; see Table~\ref{tab:GMCs} and Fig.~\ref{fig:Appendix}), the effect of differential magnification is negligible and is eventually accounted in the uncertainty on $\mu$.}; the standard deviation of the $\mu$ values within the $4\sigma$ intensity contour was used as the uncertainty on the mean magnification factor. When the $\geq 3\sigma$ intensity contours of emissions associated with different multiple images were spatially blended in a channel map, we tried at the best effort to deblend the signal by redistributing the blended flux among images proportionally to their respective fluxes measured above the threshold at which they were unblended. The lensing-corrected CO(4--3) line-integrated fluxes ($F_{\mathrm{CO(4-3)}}$) derived for the different multiple images of the same GMC were reassuring, as they agreed within less than a factor of 2 ($1-2\sigma$ flux error). We also extracted for each GMC image the integrated CO(4--3) line spectrum and performed a Gaussian function fit to the line profile, as illustrated in Fig.~\ref{fig:Appendix-spectra} for a sub-set of CO(4--3) line spectra. For the same GMC image, the corresponding flux was found in agreement, within measurement errors, with $F_{\mathrm{CO(4-3)}}$ derived from channel intensity maps. A large flux difference (close to a factor of 3) was only obtained for two GMCs with IDs 6 and 7 that we detected at 100 per cent fidelity in one channel map. In what follows we considered for these two clouds the lower $F_{\mathrm{CO(4-3)}}$ values measured from the channel maps.

Adopting the CO luminosity correction factor $r_{4,1} = L^{\prime}_{\mathrm{CO(4-3)}}/L^{\prime}_{\mathrm{CO(1-0)}} = 0.33$, as inferred for the Cosmic Snake galaxy \citep{Dessauges19}, and assuming a CO-to-H$_2$ conversion factor ($\alpha_{\mathrm{CO}}$), $M_{\mathrm{molgas}}$ of each multiple image of a given GMC was then calculated following:
\begin{equation}
M_{\mathrm{molgas}} = 
\left(\frac{\alpha_{\mathrm{CO}}}{\mathrm{M_{\odot}(K~km~s^{-1}~pc^2)^{-1}}}\right)
\left(\frac{L^{\prime}_{\mathrm{CO(4-3)}}/0.33}{\mathrm{K~km~s^{-1}~pc^2}}\right)~\mathrm{M_{\odot}},
\label{eq:Mmolgas}
\end{equation}
where the CO(4--3) luminosity ($L^{\prime}_{\mathrm{CO(4-3)}}$) was derived from $F_{\mathrm{CO(4-3)}}$ measured for that image \citep{Solomon97}.

The size of each multiple image of a given GMC was determined by first measuring in each channel map the radius of all the CO(4--3) emissions encompassed by that image.
We then adopted the radius of the emission with the highest significance level (i.e. highest S/N) among the $10.343~\mathrm{km~s^{-1}}$ channels as the radius of that image, and used the standard deviation of all measured radii, added in quadrature to the measurement error, as the error on the radius. To derive the radius, we applied the method used for the Cosmic Snake GMCs \citep{Dessauges19}. As a first step, we measured the major ($a$) and minor ($b$) semi-axes of the elliptical region that best fits the area subtended by the $4\sigma$ intensity contour of the emission associated with a given multiple image, and obtained the equivalent circularized radius, $R^{4\sigma}_{\mathrm{circ}} = \sqrt{ab}$. Considering a two-dimensional Gaussian distribution of the CO(4--3) emission, as it is generally assumed, we then derived the circularized radius at the full-width half-maximum (FWHM) as $R_{\mathrm{FWHM}} = R^{4\sigma}_{\mathrm{circ}}\sqrt{\ln(2)/\ln(F_{\mathrm{peak}}/\mathrm{RMS}/4)}$, where $F_{\mathrm{peak}}$ is the peak flux. Finally, to obtain the physical radius ($R$), we deconvolved $R_{\mathrm{FWHM}}$ from the beam and corrected the deconvolved $R_{\mathrm{FWHM}}$ from the lensing using the same $\mu$ as for the lensing-correction of the corresponding $F_{\mathrm{CO(4-3)}}$. In the case of emissions, associated with different multiple images, having their $4\sigma$ contours spatially blended in a channel map, $R_{\mathrm{FWHM}}$ was derived by using the unblended higher S/N intensity contours. 
The resulting $R$ we measured for the different multiple images of the same GMC ended up to globally agree within less than a factor of 2 ($1-2\sigma$ radius error).

The velocity dispersion of each multiple image of a given GMC was estimated by measuring in the CO(4--3) velocity dispersion second-moment map 
the mean velocity dispersion over the area subtended by the $4\sigma$ intensity contours of all the CO(4--3) emissions encompassed by that image. We adopted the largest velocity dispersion measured as the velocity dispersion ($\sigma_{v}$) of the image, and used the standard deviation of all velocity dispersions, added in quadrature to the measurement error, as the error on $\sigma_{v}$. Here again, we got an agreement between $\sigma_{v}$ derived for the different multiple images of the same GMC within less than a factor of 2 ($1-2\sigma$ velocity dispersion error). However, as these $\sigma_{v}$ were relying on the global second-moment map integrated over the whole (12-channels) A521-sys1 CO(4--3) emission, and not over the specific channels of the CO emission of the respective GMCs (which would have generated noisy second-moment maps), we also inferred the GMC $\sigma_{v}$ measurements
from the line FWHM obtained from the Gaussian fit of the integrated CO(4--3) line spectrum extracted for each GMC image (see a sub-set of those spectra and fits in Fig.~\ref{fig:Appendix-spectra}), and corrected for the final channel spacing (see the {\it ALMA} Technical Handbook). We found consistent $\sigma_{v}$, within measurement errors, for the two methods. This consistency 
suggested that the $\sigma_{v}$ measurements were not too affected by the beam smearing; as further supported by the fact that the GMCs with IDs 10 to 14, located near the galactic center where the beam smearing effect is expected to be the largest, were not found to be biased toward high $\sigma_{v}$.

The adopted physical properties --~$\mu$, $F_{\mathrm{CO(4-3)}}$, $M_{\mathrm{molgas}}$, $R$, and $\sigma_{v}$~-- of the 14 GMCs identified in the A521-sys1 galaxy are listed in Table~\ref{tab:GMCs}; we chose the measurements obtained for the less blended multiple images in channel maps of each GMC. The corresponding CO(4--3) emission line spectra are shown in Fig.~\ref{fig:Appendix-spectra}. No specific difference is found between the physical properties of the GMCs located near the galactic center and those located in the spiral arm. The two sets of GMCs encompass the same range of $M_{\mathrm{molgas}}$, $R$, and $\sigma_{v}$ values.

\begin{figure}
\includegraphics[width=0.46\textwidth,clip]{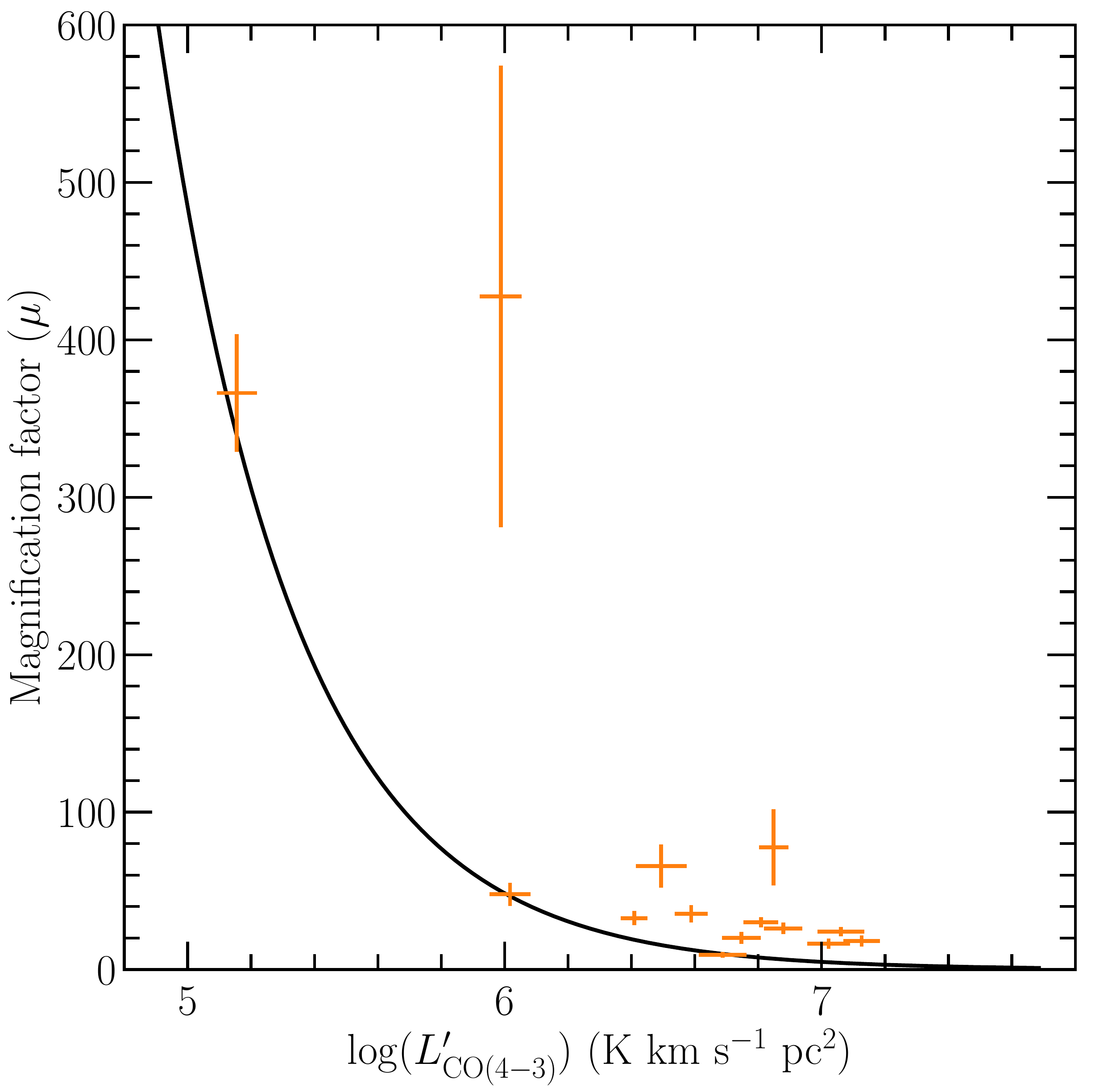}
\includegraphics[width=0.46\textwidth,clip]{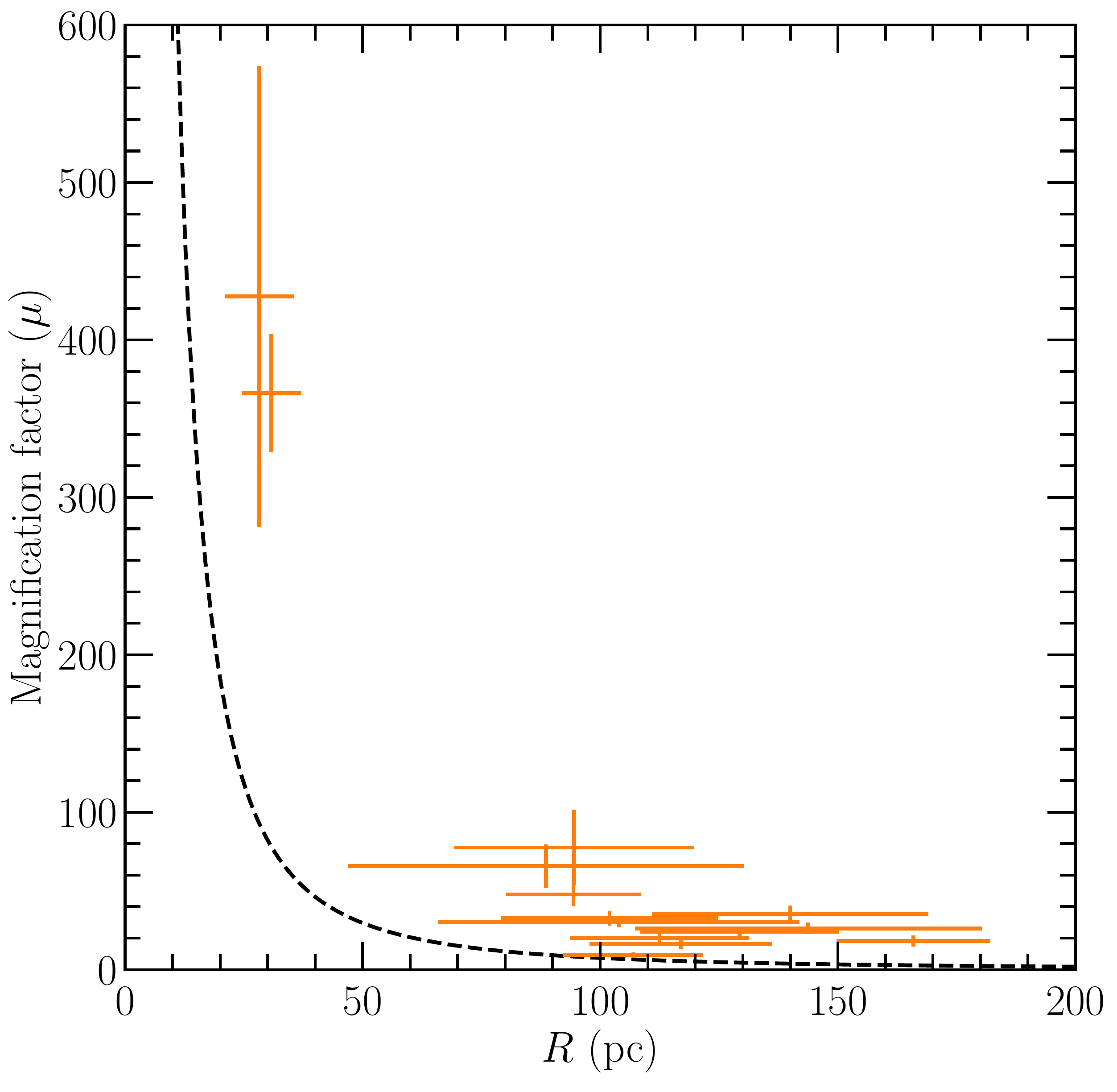}
\caption{The top and bottom panels show, respectively, the lensing-corrected CO(4--3) luminosities and beam-deconvolved circularised FWHM radii of the A521-sys1 GMCs plotted as a function of their mean lensing magnification factor. The solid line (top panel) defines the magnification-dependent $4.8\sigma$ flux detection limit, converted to the CO(4--3) luminosity limit, corresponding to the 100 per cent fidelity (see Sect.~\ref{sect:GMC-search}). The dashed line (bottom panel) defines the magnification-dependent equivalent circularised beam resolution limit. The 14 identified GMCs are spatially resolved with their $L^{\prime}_{\mathrm{CO(4-3)}}$ and $R$ lying above the respective detection limits.}
\label{fig:ampliL43-R}
\end{figure}

The 14 GMCs are spatially resolved, as their radii and masses are larger than, respectively, the magnification-dependent equivalent circularized beam resolution limit and the magnification-dependent $4.8\sigma$ flux detection limit (corresponding to the 100 per cent fidelity), as shown in Fig.~\ref{fig:ampliL43-R}. Nevertheless, we cannot exclude that some of the identified GMCs are not GMC associations, composed of several GMCs superposed along the line of sight that only observations with better spectral resolution ($\ll 10~\mathrm{km~s^{-1}}$) will resolve. Projection effects are indeed likely given the inclination of $72^{\circ}$ of the A521-sys1 galaxy \citep{Patricio18}. Therefore, we should remember that the inferred GMC masses and radii can still be upper limits.

Altogether the 14 GMCs (corrected for lensing) account for about 11 per cent of the total lensing-corrected CO(4--3) velocity-integrated flux we measured in the A521-sys1 arc, where the GMCs were searched for and identified; about 55 per cent of the A521-sys1 source galaxy is imaged in the arc. This means there still is a substantial amount of CO(4--3) molecular gas in the galaxy beside the detected GMCs, which is an unresolved `background'. This gas can be in a diffuse phase, but 
we consider it is mostly made of sub-clouds, which are not resolved due to our limited angular resolution with {\it ALMA}. A hierarchical fragmentation over several orders of magnitude and well below $\sim 10$~pc is indeed expected, and is believed to result from the hierarchical nature of turbulence \citep{Elmegreen97,Padoan02,Hopkins13a,Hopkins13b}. As a result, the molecular gas follows a quasi-hierarchical structure, containing sub-clouds, clouds, and sub-clouds inside clouds. Therefore, in the same way as proceeded by \citet{Solomon87} for local GMCs where they took all the mass above a certain level of integrated intensity, no background was subtracted when we measured the mass of GMCs in order to take all the mass within the boundary of a GMC including all sub-clumps.

Finally, we would like to stress that we did not find much smooth emission in our high-resolution {\it ALMA} data of A521-sys1. This is not due to the filtered out large-scale CO(4--3) emission, since the total integrated {\it ALMA} CO(4--3) emission flux\footnote{We integrated the {\it ALMA} CO(4--3) emission over all A521-sys1 multiple images, since they were all covered by the large {\it IRAM} 30~m beam.} agrees with the fluxes inferred from the CO(2--1) and CO(3--2) lines detected with the {\it IRAM} 30~m single dish antenna (F.~Boone, private communication),
for a CO spectral line energy distribution typical of galaxies at $z\sim 1-2$ \citep{Daddi15,Dessauges15}.

\begin{figure}
\includegraphics[width=0.47\textwidth,clip]{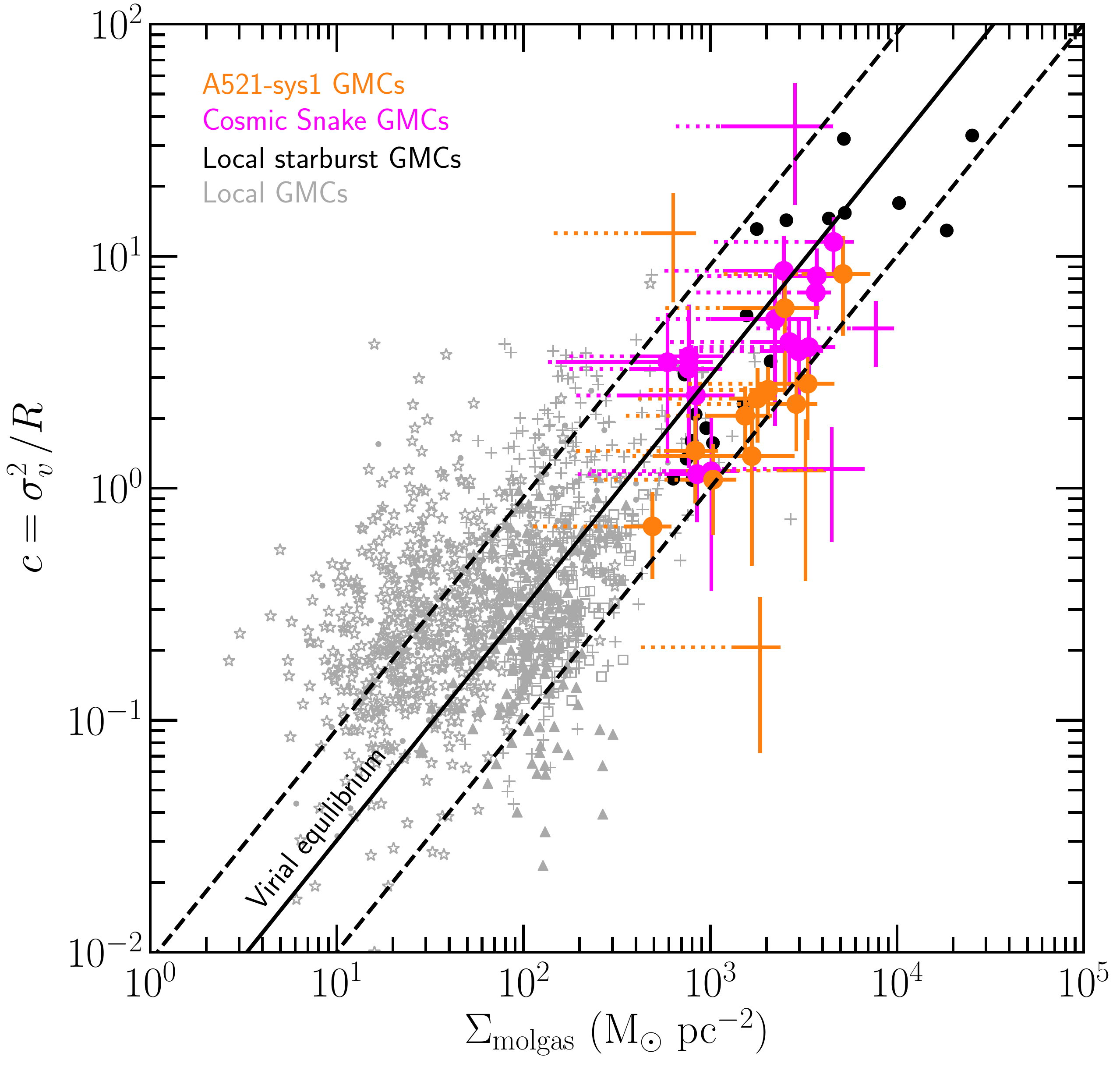}
\caption{Size-velocity dispersion coefficients plotted as a function of molecular gas mass surface density. The solid line shows the expected one-to-one relation for virialized GMCs, $\Sigma_{\mathrm{molgas}}\approx 330 \sigma_{v}^2/R$, with the dashed lines marking the mean spread ($\sim 0.4-0.5~\mathrm{dex}$) of GMCs hosted in local quiescent galaxies (grey dots \citep{Bolatto08}, triangles \citep{Heyer09}, squares \citep{Donovan13}, crosses \citep{Colombo14}, and stars \citep{Corbelli17}) and nearby starbursting/merger galaxies (black filled circles \citep{Wei12,Leroy15}). The orange data points correspond to the A521-sys1 GMCs with the dotted lines showing the range of possible $\Sigma_{\mathrm{molgas}}$ determined with two different $\alpha_{\mathrm{CO}}$, the Milky Way and starburst values of $4.36$ and $1.0~\mathrm{M_{\odot}(K~km~s^{-1}~pc^2)^{-1}}$, respectively. Eleven out of the 14 A521-sys1 GMCs (orange filled circles) are virialized, since located within a factor of 3 from the one-to-one $C$--$\Sigma_{\mathrm{molgas}}$ relation (dashed lines), independently of $\alpha_{\mathrm{CO}}$. For comparison, we also show the 17 GMCs identified in the Cosmic Snake galaxy (magenta data points), among which 14 are found to be virialized (magenta filled circles \citep{Dessauges19}).}
\label{fig:virial-equilibrium}
\end{figure}

\subsection{Virial equilibrium and CO-to-H$_2$ conversion factor}
\label{sect:virial}

A way to give credit to the identified GMCs in the A521-sys1 galaxy is to determine whether they are gravitationally bound structures in virial equilibrium. Most of the GMCs in local/nearby galaxies, as well as in the Cosmic Snake galaxy at $z\simeq 1$, are indeed virialized 
\citep{Heyer09,Sun18,Dessauges19,Rosolowsky21}. A way to determine the dynamical state of GMCs is to consider the size-velocity dispersion coefficient ($C = \sigma_{v}^2/R$) that is linearly proportional to 
the gas mass surface density ($\Sigma_{\mathrm{molgas}}$) for a virialized GMC. In Fig.~\ref{fig:virial-equilibrium} we show that within a factor of 3 11 out of the 14 A521-sys1 GMCs are distributed around the one-to-one $C$--$\Sigma_{\mathrm{molgas}}$ relation, independently of the $\alpha_{\mathrm{CO}}$ considered between the Milky Way value of $4.36~\mathrm{M_{\odot}(K~km~s^{-1}~pc^2)^{-1}}$ and the starburst value of $1~\mathrm{M_{\odot}(K~km~s^{-1}~pc^2)^{-1}}$ \citep{Bolatto13}. The factor of 3 uncertainty in this parameter space corresponds to the standard deviation of the local GMC distribution ($\sim 0.4-0.5$~dex) and to measurement errors. Therefore, within this uncertainty GMCs are usually assumed to be in virial equilibrium, although some of them can be transient structures or collapsing. 

\begin{figure}
\includegraphics[width=0.462\textwidth,clip]{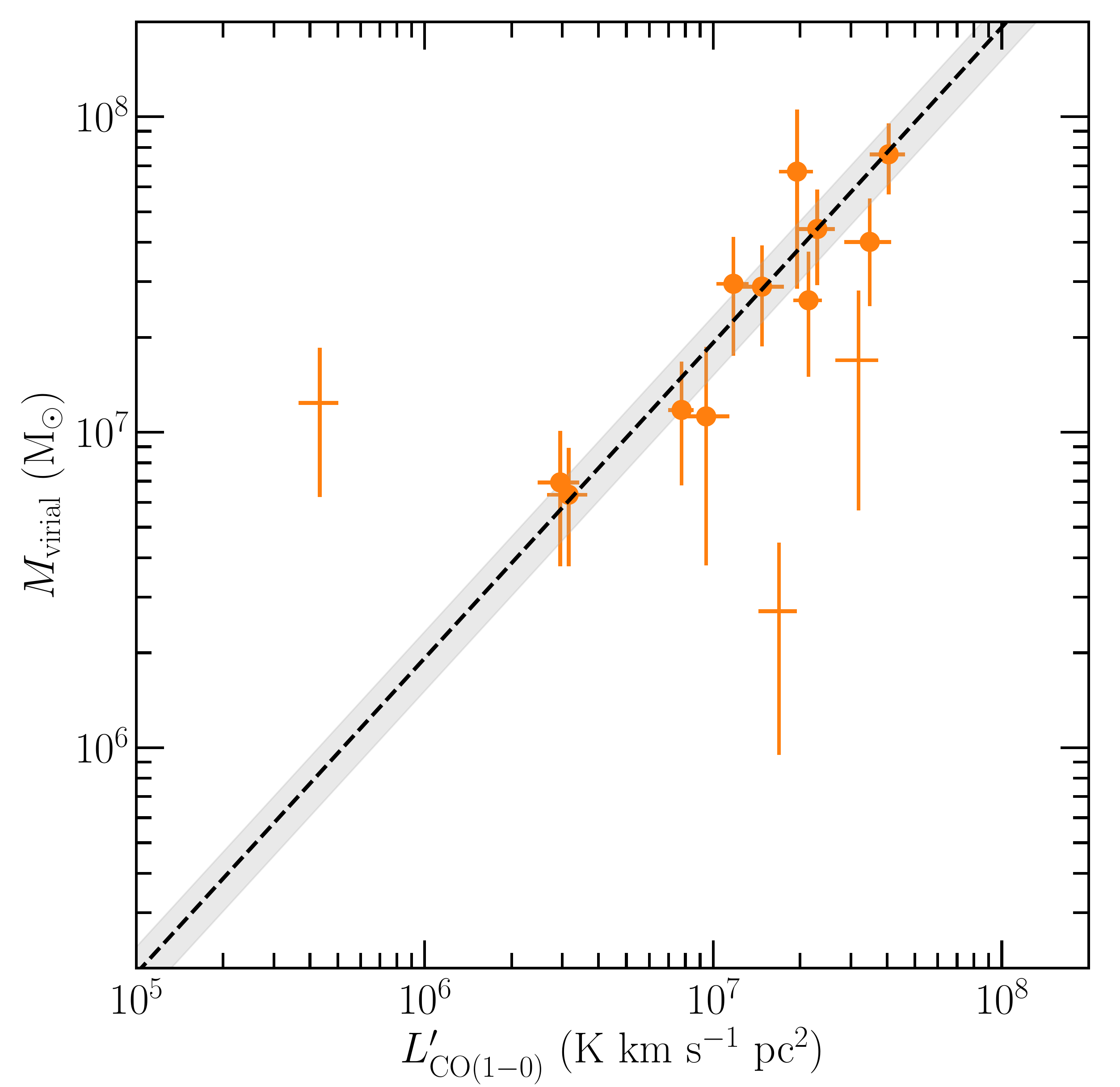}
\caption{Virial masses of the A521-sys1 GMCs plotted as a function of their lensing-corrected CO(1--0) luminosity (derived from the measured CO(4--3) luminosity assuming $r_{4,1} = 0.33$). The orange filled circles correspond to the 11 GMCs found in virial equilibrium (Fig.~\ref{fig:virial-equilibrium}). They are distributed about the mean CO-to-H$_2$ conversion factor of $\alpha_{\mathrm{CO}} = 1.9~\mathrm{M_{\odot}(K~km~s^{-1}~pc^2)^{-1}}$ (dashed line) with a standard deviation of $0.4~\mathrm{M_{\odot}(K~km~s^{-1}~pc^2)^{-1}}$ (grey shaded area).}
\label{fig:virial-LCO}
\end{figure}

For virialized GMCs, their virial mass is expected to be linearly proportional to CO luminosity \citep{Bolatto08,Bolatto13}. Therefore, we can obtain $\alpha_{\mathrm{CO}}$ measurements from the GMC internal kinematics. As done for the Cosmic Snake GMCs \citep{Dessauges19}, to compute the virial mass ($M_{\mathrm{virial}}$) we assumed GMCs to be spherical and have their density profile inversely proportional to radius \citep{Solomon87,Hughes13}:
\begin{equation}
M_{\mathrm{virial}} = 1040\left(\frac{\sigma_{v}}{\mathrm{km~s^{-1}}}\right)^2 \left(\frac{R}{\mathrm{pc}}\right)~\mathrm{M_{\odot}} .
\label{eq:Mvirial}
\end{equation}
In Fig.~\ref{fig:virial-LCO} we show $M_{\mathrm{virial}}$ as a function of $L^{\prime}_{\mathrm{CO(1-0)}}$ (lensing-corrected) of the 14 A521-sys1 GMCs at $z\simeq 1$. The 11 GMCs, found in virial equilibrium (Fig.~\ref{fig:virial-equilibrium}), are distributed about the mean $\alpha_{\mathrm{CO}} = 1.9~\mathrm{M_{\odot}(K~km~s^{-1}~pc^2)^{-1}}$ with a standard deviation of $0.4~\mathrm{M_{\odot}(K~km~s^{-1}~pc^2)^{-1}}$. This suggests a lower $\alpha_{\mathrm{CO}}$ than in the Cosmic Snake galaxy, where we determined, also from the GMC internal kinematics, $\alpha_{\mathrm{CO}} = 3.8\pm 1.1~\mathrm{M_{\odot}(K~km~s^{-1}~pc^2)^{-1}}$ \citep{Dessauges19}. Both galaxies at $z\simeq 1$ yet have a solar metallicity \citep{Patricio18} and a comparable SFR of $\sim 26-30~\mathrm{M_{\odot}~yr^{-1}}$ \citep{Nagy22}, what changes is the median $\Sigma_{\mathrm{molgas}}$ of their respective GMCs. The significantly higher SFR of these $z\simeq 1$ galaxies than of the Milky Way galaxy implies a stronger photodissociating radiation, and explains their globally lower $\alpha_{\mathrm{CO}}$ with respect to the Milky Way value.
However, in the case of the Cosmic Snake galaxy the GMC $\Sigma_{\mathrm{molgas}}$  happen to be so high, with a median value of $\sim 2300~\mathrm{M_{\odot}~pc^{-2}}$ (for the above cited $\alpha_{\mathrm{CO}}$), that the GMCs are shielded from the ambient radiation. This is less the case for the A521-sys1 GMCs, which have a median $\Sigma_{\mathrm{molgas}} \sim 800~\mathrm{M_{\odot}~pc^{-2}}$ (for the derived $\alpha_{\mathrm{CO}}$), and therefore the CO-to-H$_2$ conversion factor in A521-sys1 could have ended up to be lower. However, globally the factor of two difference in the $\alpha_{\mathrm{CO}}$ values derived for the two galaxies at $z\simeq 1$, both with about solar metallicities, is within the spread of the $\alpha_{\mathrm{CO}}$ measurements obtained for nearby solar metallicity galaxies \citep[e.g.][]{Leroy11,Sandstrom13}.


\section{Discussion}
\label{sect:discussion}

\subsection{Larson scaling relations}
\label{sect:Larson-relations}

The GMCs hosted in the A521-sys1 galaxy have high $M_{\mathrm{molgas}} = 8.2\times 10^5-7.7\times 10^7~\mathrm{M}_{\odot}$\footnote{Obtained by adopting $\alpha_{\mathrm{CO}} = 1.9\pm 0.4~\mathrm{M_{\odot}(K~km~s^{-1}~pc^2)^{-1}}$ derived in Sect.~\ref{sect:virial}.}, with a median of $3.0\times 10^7~\mathrm{M}_{\odot}$. This is two orders of magnitude higher than the median $M_{\mathrm{molgas}}$ of GMCs found in the Milky Way, local dwarf galaxies, and nearby spirals \citep[$2.9\times 10^5~\mathrm{M}_{\odot}$;][]{Bolatto08,Heyer09,Donovan13,Colombo14,Corbelli17}. With their $R = 28-166~\mathrm{pc}$, they also have a high median $\Sigma_{\mathrm{molgas}}$ (see Sect.~\ref{sect:virial}), almost 10 times higher than local GMCs ($\Sigma_{\mathrm{molgas}}\sim 100~\mathrm{M_{\odot}~pc^{-2}}$). Their $\sigma_{v} = 5-25~\mathrm{km~s^{-1}}$ are large enough to enable them to be in virial equilibrium. 

\begin{figure}
\includegraphics[width=0.455\textwidth,clip]{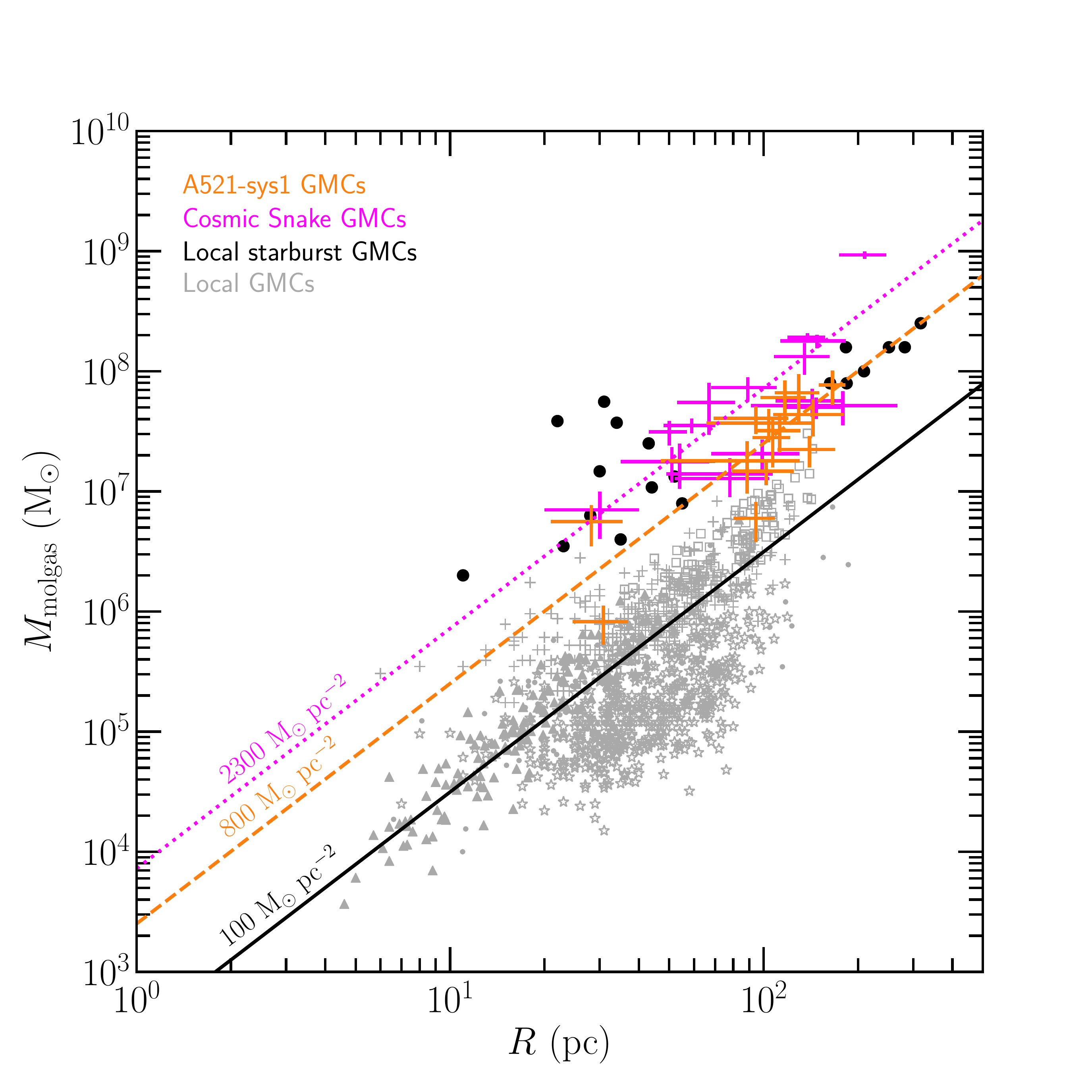}
\includegraphics[width=0.455\textwidth,clip]{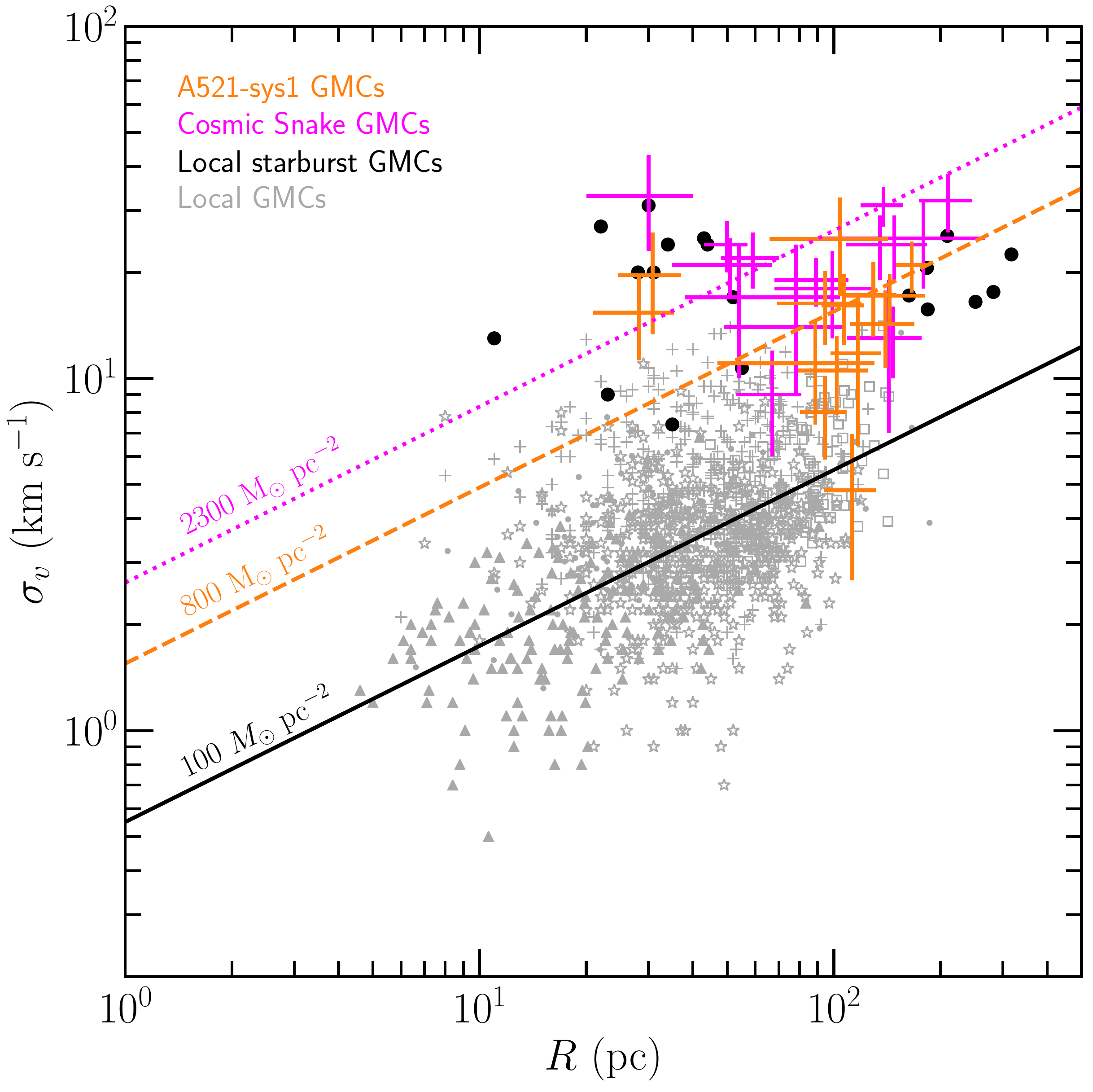}
\caption{The top panel shows the first Larson scaling relation with the molecular gas masses plotted as a function of radius for the GMCs hosted in the A521-sys1 galaxy at $z=1.043$ (orange data points), the Cosmic Snake galaxy at $z=1.036$ (magenta data points \citep{Dessauges19}), local quiescent galaxies (grey dots \citep{Bolatto08}, triangles \citep{Heyer09}, squares \citep{Donovan13}, crosses \citep{Colombo14}, and stars \citep{Corbelli17}) and nearby starbursting/merger galaxies (black filled circles \citep{Wei12,Leroy15}). The lines correspond to the median molecular gas mass surface densities of $100~\mathrm{M_{\odot}~pc^{-2}}$ (solid line), $800~\mathrm{M_{\odot}~pc^{-2}}$ (orange dashed line), and $2300~\mathrm{M_{\odot}~pc^{-2}}$ (magenta dotted line) as measured for, respectively, the local GMCs, the A521-sys1 GMCs, and the Cosmic Snake GMCs. The GMCs at $z\simeq 1$ are clearly much denser than typical local GMCs, but have comparable $\Sigma_{\mathrm{molgas}}$ to those of GMCs hosted in nearby starbursting/merger galaxies. 
The bottom panel shows the second Larson scaling relation with the internal velocity dispersions plotted as a function of radius for the same GMC populations as in the top panel. We plot the $\sigma_{v}\propto R^{0.5}$ relation for the three median $\Sigma_{\mathrm{molgas}}$ values derived in the top panel.
The internal velocity dispersions of $z\simeq 1$ GMCs are on average larger than those of typical local GMCs to enable them to remain in virial equilibrium despite their higher $\Sigma_{\mathrm{molgas}}$. 
The adopted $\alpha_{\mathrm{CO}}$ for the A521-sys1 and Cosmic Snake GMCs are given in Sect.~\ref{sect:virial}.}
\label{fig:Mmolgas-v-R}
\end{figure}

These measurements, together with the physical properties derived for the Cosmic Snake GMCs, confirm that GMCs hosted in main-sequence galaxies at $z\simeq 1$ differ from GMCs in the local Universe. This is very well assessed when considering the Larson scaling relations, used as the benchmark for the typical local GMC population \citep{Larson81,Bolatto08}. In Fig.~\ref{fig:Mmolgas-v-R} we plot, respectively, the first Larson scaling relation $M_{\mathrm{molgas}}$--$R$  
that shows a constant $\Sigma_{\mathrm{molgas}}$ at $\sim 100~\mathrm{M_{\odot}~pc^{-2}}$ for local GMCs (grey data points; top panel), and the second Larson scaling relation $\sigma_{v}$--$R$ 
that calibrates $\Sigma_{\mathrm{molgas}}$ of virialized local GMCs (grey data points; bottom panel). The GMCs at $z\simeq 1$ are clearly offset from these two relations. On the other hand, as already suggested in \citet{Dessauges19}, the high-redshift GMCs very much resemble GMCs hosted in the Antennae galaxy merger \citep{Wei12} and the NGC~253 nuclear starburst \citep{Leroy15}, which are also offset from the Larson relations because immersed in an ISM characterized by an enhanced SFR per unit mass of molecular gas and a strong ambient pressure \citep[see also the GMCs hosted in the starburst galaxy NGC~4826;][]{Rosolowsky05,Rosolowsky21}. It has now been evidenced by a number of studies of GMCs in nearby galaxies that the physical properties of GMCs are indeed dependent on their surrounding local environment \citep[e.g.][]{Hughes13,Sun18,Sun20}. The A521-sys1 GMCs further support the strong influence of the ambient ISM, which at $z\simeq 1$ clearly differs from the ISM of nearby quiescent galaxies as discussed in the next Sect.~\ref{sect:pressure}.


\subsection{Pressure equilibrium}
\label{sect:pressure}

\begin{figure}
\includegraphics[width=0.47\textwidth,clip]{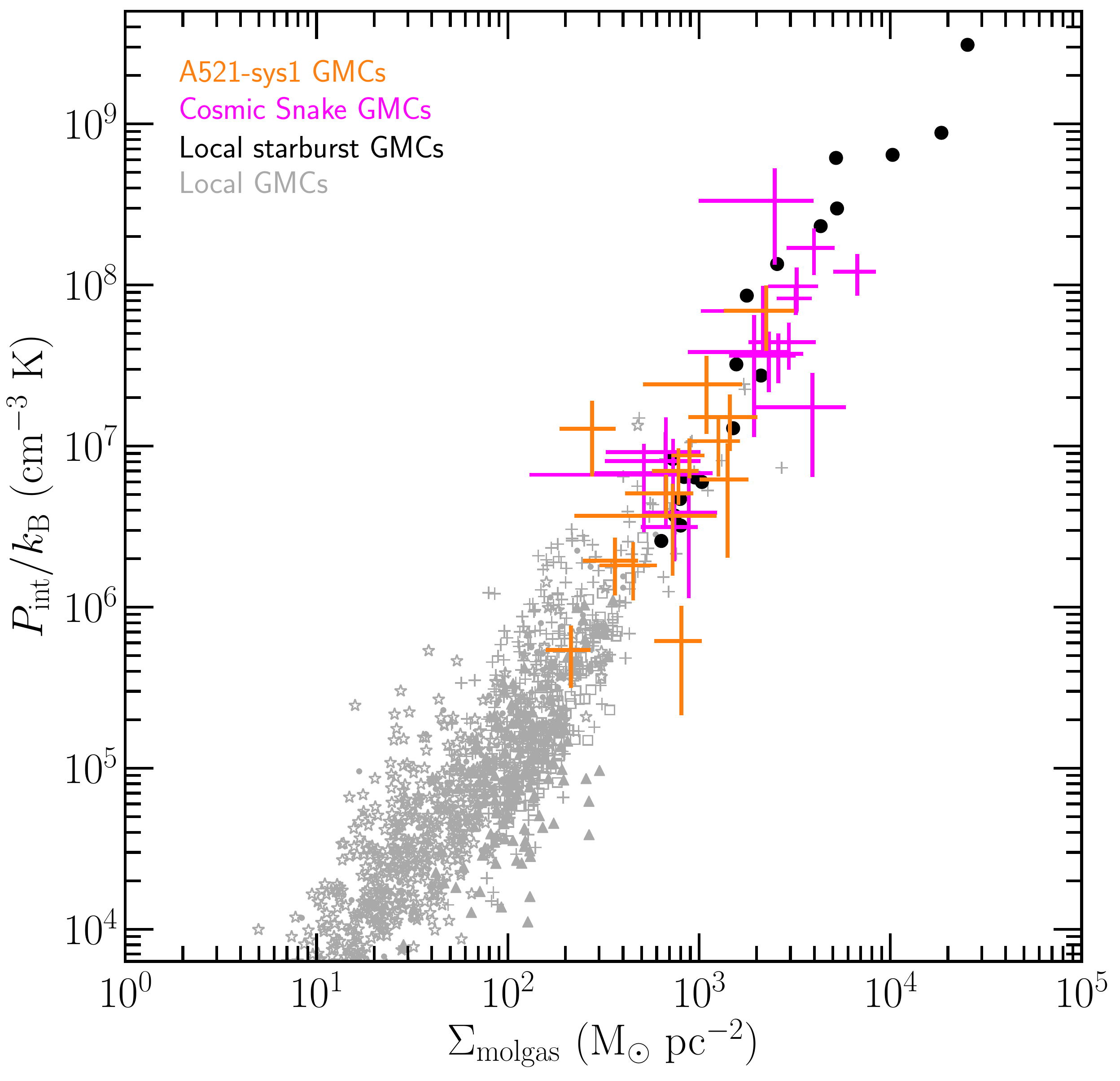}
\caption{Internal kinetic pressures plotted as a function of molecular gas mass surface density for the same GMC populations as in Fig.~\ref{fig:Mmolgas-v-R}. The GMCs at $z\simeq 1$ (orange and magenta data points) extend the $P_{\mathrm{int}}/k_{\mathrm{B}}$--$\Sigma_{\mathrm{molgas}}$ correlation of typical local GMCs over three orders of magnitude in $\Sigma_{\mathrm{molgas}}$.
The adopted $\alpha_{\mathrm{CO}}$ are given in Sect.~\ref{sect:virial}.}
\label{fig:pressure}
\end{figure}

The internal kinetic pressure ($P_{\mathrm{int}}/k_{\mathrm{B}}$) of the A521-sys1 GMCs can be determined following \citet{Hughes13}:
\begin{equation}
\frac{P_{\mathrm{int}}}{k_{\mathrm{B}}} = 1176 \left(\frac{M_{\mathrm{molgas}}}{\mathrm{M}_{\odot}}\right) \left(\frac{R}{{\mathrm{pc}}}\right)^{-3} \left(\frac{\sigma_{v}}{{\mathrm{km~s^{-1}}}}\right)^2~\mathrm{cm^{-3}~K}.
\label{eq:int-pressure}
\end{equation}
As shown in Fig.~\ref{fig:pressure}, we found strong $P_{\mathrm{int}}/k_{\mathrm{B}}$ between $10^{5.7}-10^{7.8}~\mathrm{cm^{-3}~K}$, comparable to the distribution of the values determined for the Cosmic Snake GMCs at their low-pressure end. The $P_{\mathrm{int}}/k_{\mathrm{B}}$ measurements in the $z\simeq 1$ GMCs show a dependence on $\Sigma_{\mathrm{molgas}}$, which very well extends, toward higher $\Sigma_{\mathrm{molgas}}$, the $P_{\mathrm{int}}/k_{\mathrm{B}}$--$\Sigma_{\mathrm{molgas}}$ correlation reported for typical local GMCs \citep{Hughes13,Sun18}.

In \citet{Swinbank15} and \citet{Dessauges19} it was argued that the strong $P_{\mathrm{int}}/k_{\mathrm{B}}$ of high-redshift GMCs is linked to the strong hydrostatic pressure at the disc midplane ($P_{\mathrm{ext}}/k_{\mathrm{B}}$) observed in their host galaxies, which was suggested to approximate the pressure at the boundary of molecular clouds \citep{Elmegreen89,Blitz04}. For a two-component disc of gas and stars, $P_{\mathrm{ext}}/k_{\mathrm{B}}$ can be estimated as:
\begin{equation}
\frac{P_{\mathrm{ext}}}{k_{\mathrm{B}}} \approx \frac{\pi G}{2} \Sigma_{\mathrm{gas}} \left(\Sigma_{\mathrm{gas}}+\frac{\sigma_{\mathrm{gas}}}{\sigma_{\mathrm{stars}}}\Sigma_{\mathrm{stars}}\right)~\mathrm{cm^{-3}~K},
\label{eq:hydrostatic-pressure}
\end{equation}
where $G$ is the gravitational constant, and $\Sigma_{\mathrm{gas}}$, $\Sigma_{\mathrm{stars}}$, and $\sigma_{\mathrm{gas}}$, $\sigma_{\mathrm{stars}}$, respectively, are the neutral (atomic + molecular) gas and stellar mass surface densities and velocity dispersions \citep{Elmegreen89,Blitz04}. Similarly to \citet{Dessauges19} and \citet{Swinbank15}, we assumed the neutral gas of high-redshift galaxies to be dominated by the molecular gas, and $\sigma_{\mathrm{gas}}\approx \sigma_{\mathrm{stars}}$. We derived $\Sigma_{\mathrm{gas}}$ and $\Sigma_{\mathrm{stars}}$ from the total molecular gas and stellar masses, respectively, measured in the uniformly lensed image of the A521-sys1 galaxy (see Fig.~\ref{fig:HST-ALMA}). This gave us $P_{\mathrm{ext}}/k_{\mathrm{B}}\sim 10^{7.2}~\mathrm{cm^{-3}~K}$, namely a factor of $\sim 7$ weaker than in the Cosmic Snake galaxy \citep{Dessauges19}, but a factor of $\sim 100-1000$ stronger than the hydrostatic pressure of nearby quiescent galaxies, including the Milky Way disc \citep{Hughes13}. 

The results obtained for the A521-sys1 galaxy, showing a weaker $P_{\mathrm{ext}}/k_{\mathrm{B}}$ \citep[mostly because of the lower molecular gas mass surface density of A521-sys1;][]{Nagy22} together with slightly more moderate GMC physical properties (density, turbulence, internal pressure) than in the Cosmic Snake galaxy (see Figs.~\ref{fig:Mmolgas-v-R} and \ref{fig:pressure}), bring an additional evidence that GMCs form with properties adjusted to the ambient ISM conditions particular to the host galaxy, and this not only in local galaxies but also in high-redshift galaxies. In summary, the GMCs of main-sequence galaxies at $z\simeq 1$ show variations in their properties from galaxy to galaxy. Their properties resemble those of local GMCs hosted in starbursting media, where the ambient pressure is comparable to the one observed in the ISM of high-redshift galaxies \citep{Rosolowsky05,Hughes13,Leroy15,Sun20}. However, they are different from the properties of typical local GMCs, because the strong $P_{\mathrm{ext}}/k_{\mathrm{B}}$ of distant galaxies exceeds the self-gravity of these local GMCs that would be rapidly compressed.


\subsection{Efficiency of star formation}
\label{sect:SFE}

The Cosmic Snake GMCs at $z\simeq 1$ were found to be highly supersonic, with a median Mach number ($\mathcal{M}$) of $80\pm 13$ \citep{Dessauges19}, almost 10 times higher than that of typical local GMCs \citep{Leroy15}. We suggested that such a highly supersonic turbulence may imply a particularly efficient formation of stars in these GMCs, given that star clusters form by gravitational collapse of shock-compressed density fluctuations generated from the supersonic turbulence in GMCs \citep{McKee07,Brunt09}. And, indeed, we estimated the GMC star-formation efficiency ($\epsilon$) to be as high as $\sim 26-34$ per cent from the simple comparison of the mass distribution of GMCs and stellar clumps identified in the Cosmic Snake galaxy \citep{Dessauges19}. This is sensibly higher than $\epsilon\lesssim 6$\% reported in nearby galaxies \citep{Evans09,Chevance20,Kim21}, but consistent with the $\epsilon$--$\Sigma_{\mathrm{molgas}}$ scaling-relation predicted for molecular clouds by numerical simulations \citep{Grudic18,Grudic22}.

The A521-sys1 GMCs discussed here and the stellar clumps analysed by \citet{Messa22} offer a new opportunity to investigate $\epsilon$ in a galaxy at $z\simeq 1$. We derived the GMC Mach number ($\mathcal{M}$) following the prescriptions of \citet{Leroy15}:
\begin{equation}
\mathcal{M} = \frac{\sqrt{3} \sigma_{v}}{c_{\mathrm{s}}},
\label{eq:Mach-number}
\end{equation}
where $c_{\mathrm{s}} = 0.45~\mathrm{km~s^{-1}}$ is the sound speed estimated for molecular hydrogen at $20-50$~K. We derived a high median $\mathcal{M} = 60\pm 11$ for the 14 A521-sys1 GMCs, showing they are
highly supersonic, as are the Cosmic Snake GMCs, and may hence potentially experience a high $\epsilon$. 

\begin{figure}
\includegraphics[width=0.442\textwidth,clip]{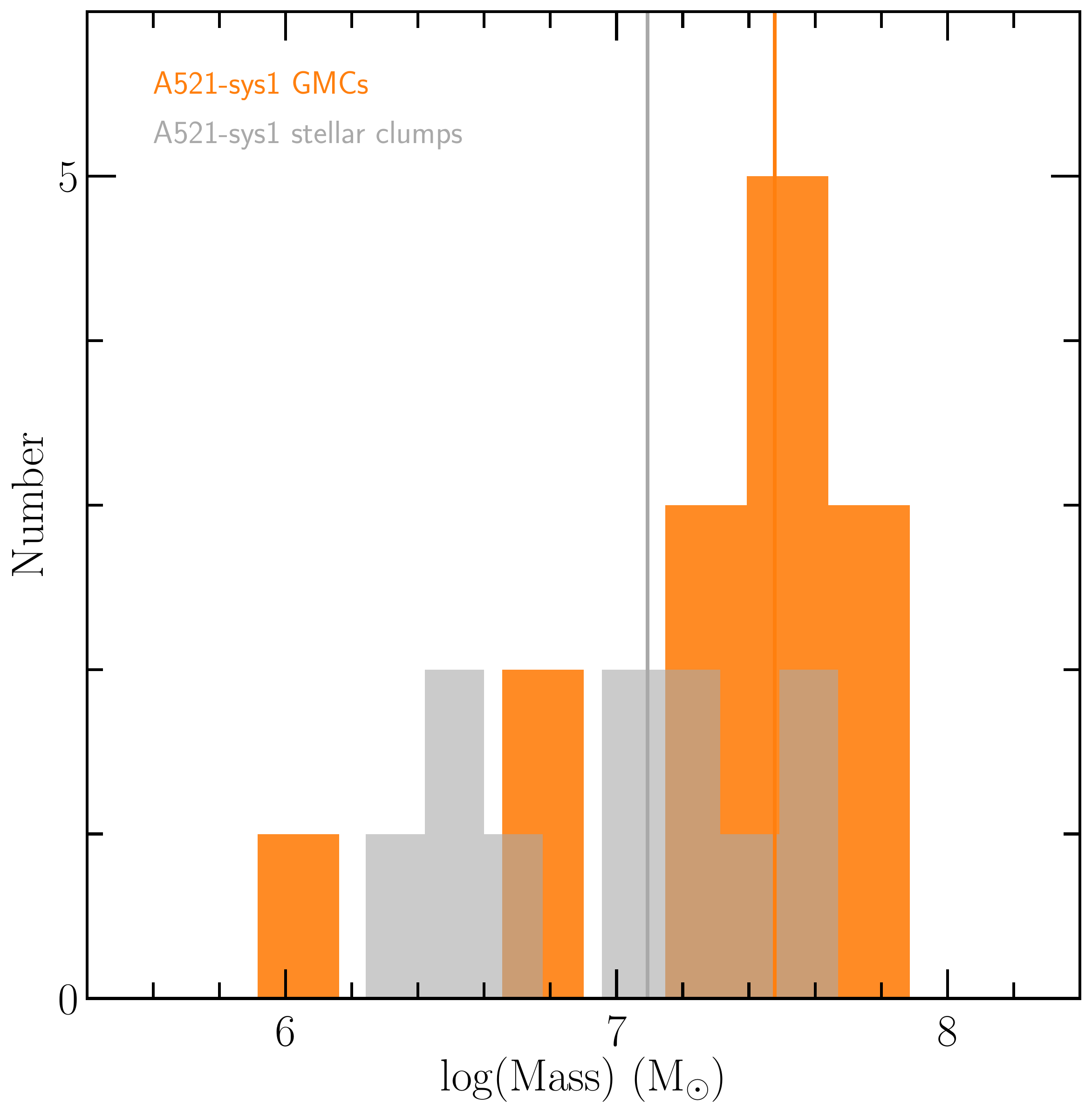}
\caption{Molecular gas mass distribution of 14 GMCs (orange histogram) versus stellar mass distribution of 11 stellar clumps (grey histogram; \citet{Messa22}) hosted in the A521-sys1 galaxy and all imaged over the arc within regions where the magnification factor is larger than 4. The solid vertical lines show the respective medians of $3.0\times 10^7~\mathrm{M}_{\odot}$ for the GMCs and $1.2\times 10^7~\mathrm{M}_{\odot}$ for the stellar clumps. Their comparison suggests a fairly high efficiency of star formation as discussed in Sect.~\ref{sect:SFE}.
The adopted $\alpha_{\mathrm{CO}}$ is derived in Sect.~\ref{sect:virial}.}
\label{fig:MassGMCs-MassClumps}
\end{figure}

In Fig.~\ref{fig:MassGMCs-MassClumps} we compare the distributions of masses of the 14 GMCs and the 11 stellar clumps \citep{Messa22} identified in the A521-sys1 galaxy, all imaged over the arc within regions where $\mu >4$\footnote{We focussed only on stellar clumps imaged over the arc and that benefit from high $\mu >4$ \citep{Messa22}.}. 
This comparison can be used to estimate $\epsilon$ in the same way as for local galaxies via $\epsilon = M_{\mathrm{stars}}/(M_{\mathrm{molgas}} + M_{\mathrm{stars}})$ \citep{Evans09} given that: the identified stellar clumps are young, with ages mainly $<100$~Myr, i.e.\ likely tracing the recent star-formation episodes as do the GMCs \citep{Messa22}; and the stellar clumps and GMCs were identified at a similar spatial resolution, throughout, respectively, the {\it HST} near-UV to near-IR images \citep[with the PSF varying from 0.10 to 0.24 arcsec;][]{Messa22} and the {\it ALMA} images (with the synthesized circularized beam of 0.17 arcsec).
It appears that the most massive GMCs are only slightly more massive than the most massive stellar clumps, which suggests a relatively high $\epsilon$. Considering the median values of the molecular gas masses of the GMCs and of the stellar masses of the stellar clumps, we get $\epsilon\sim 29$ per cent in this $z\simeq 1$ galaxy, provided that the observed massive stellar clumps (star cluster complexes/associations) were formed within GMCs with comparable physical properties to those we identified.

We also estimated the star-formation efficiency per free-fall time ($\epsilon_{\mathrm{ff}}$):
\begin{equation}
\epsilon_{\mathrm{ff}} = \frac{\mathrm{SFR}}{M_{\mathrm{molgas}}} t_{\mathrm{ff}}.
\label{eq:epsilon-tff}
\end{equation}
$t_{\mathrm{ff}}$ is the free-fall time defined as $t_{\mathrm{ff}} = \sqrt{3\pi/(32 G \rho)}$, where $\rho$ is the volume gas density, and $t_{\mathrm{ff}}$ is considered as the characteristic time-scale for the GMC gravitational free-fall collapse inducing the formation of stars
\citep{Krumholz12,Federrath12}. A theoretical value of $\epsilon_{\mathrm{ff}}\sim 1$ per cent is commonly accepted in the literature (especially in numerical simulations), meaning that each free-fall time a galaxy typically converts about 1 per cent of its dense gas into stars \citep[e.g.][]{Krumholz07,McKee07,Krumholz12}. This value is globally supported by observations, although important $\epsilon_{\mathrm{ff}}$ deviations exist among different environments of nearby galaxies \citep[e.g.][]{Leroy17,Ochsendorf17,Utomo18,Schruba19,Kim21}. 

Under the simple assumption that the detected high-redshift A521-sys1 GMCs are spherical (see also Sect.~\ref{sect:virial}), their $t_{\mathrm{ff}}$ range between 1 and 6~Myr, with a median value of $\sim 3$~Myr, comparable to the free-fall times measured for GMCs in nearby galaxies (see references above). We carried out the estimation of $\epsilon_{\mathrm{ff}}$ in the A521-sys1 galaxy for three cases: (1)~considering the total $\mathrm{SFR} = 26\pm 5~\mathrm{M}_{\odot}~\mathrm{yr}^{-1}$ and the total $M_{\mathrm{molgas}} = (1.2\pm 0.2)\times 10^{10}~\mathrm{M}_{\odot}$\footnote{Derived from the CO(4--3) line emission, and calculated by adopting $r_{4,1} = 0.33$ and $\alpha_{\mathrm{CO}} = 1.9\pm 0.4~\mathrm{M_{\odot}(K~km~s^{-1}~pc^2)^{-1}}$ determined in Sect.~\ref{sect:virial}. As mentioned in Sect.~\ref{sect:GMC-properties}, the CO(2--1) and CO(3--2) emission lines, detected with the {\it IRAM} 30~m single dish antenna, give comparable total molecular gas mass estimates (F. Boone, private communication).} integrated over the entire galaxy; (2)~considering the radial $\Sigma_\mathrm{SFR}/\Sigma_{\mathrm{molgas}}$ measurements from \citet{Nagy22}, inferred in concentric elliptical rings centred on the galaxy center; and (3)~considering the total galactic $\mathrm{SFR}$, but 
the $M_{\mathrm{molgas}}$ contribution from the 14 GMCs only that means considering only the compact/dense CO gas that drives most of the star formation of the galaxy. 
For case~(1) we found a low $\epsilon_{\mathrm{ff}}\sim 0.8$ per cent, and for case~(2) $\epsilon_{\mathrm{ff}}$ ranged between 0.9 and 4.3 per cent. For case~(3) we obtained a much higher $\epsilon_{\mathrm{ff}}\sim 11$ per cent, which is more in line with the $\epsilon$ estimation above. While the $\epsilon_{\mathrm{ff}}$ derived for cases~(1) and (2) are in line with most of the measurements reported for nearby galaxies \citep[][and references above]{Utomo18}, $\epsilon_{\mathrm{ff}}$ for case~(3) stands out of the range. The difference between cases~(1)~+~(2) and case~(3) comes from the fact that the compact/dense molecular gas found in GMCs represents a small fraction of about 11 per cent in mass of the total molecular gas that is observed in the A521-sys1 galaxy (Sect.~\ref{sect:GMC-properties}). In the definition of $\epsilon_{\mathrm{ff}}$ (equation~(\ref{eq:epsilon-tff})), the molecular gas mass precisely refers to the dense gas mass that is available for the star formation \citep{Kim21}. Therefore, we believe that our estimate of $\epsilon_{\mathrm{ff}}\sim 11$ per cent for case~(3) is most likely characterising the more efficient star-formation process taking place in the $z\sim 1$ A521-sys1 galaxy compared to nearby galaxies.


\section{Conclusions}
\label{sect:conclusions}

We present the analysis of the CO(4--3) molecular gas emission of the A521-sys1 main-sequence star-forming galaxy at $z=1.043$. In the {\it HST} rest-frame UV to optical images, the galaxy is clumpy, showing  several UV-bright clumps \citep{Messa22}, and has a nice spiral structure, also traced by the CO(4--3) emission (see Fig.~\ref{fig:HST-ALMA}) that even exhibits a surface over-density of molecular gas mass in the galaxy radial profile at the location of the spiral arms \citep{Nagy22}. The kinematics of both ionized ([O\,{\sc ii}] emission) and molecular (CO(4--3) emission) gas reveals a rotating disc, with an inclination-corrected rotation velocity of $\upsilon_{\mathrm{rot}} = 130\pm 5~\mathrm{km~s^{-1}}$ \citep{Patricio18,Girard19}. The respective velocity dispersions significantly differ between $\sigma_0 = 54\pm 11~\mathrm{km~s^{-1}}$ for the ionized gas and $11\pm 8~\mathrm{km~s^{-1}}$ for the molecular gas \citep{Girard19}. We can determine the Toomre stability parameter \citep{Toomre64} for a constant rotation velocity as follows:
\begin{equation}
Q_{\mathrm{gas}} = \sqrt{2} \frac{\sigma_0}{\upsilon_{\mathrm{rot}} f_{\mathrm{gas}}},
\end{equation}
where $f_{\mathrm{gas}}$ is the molecular gas mass to total (including gas, stars, and dark matter) mass fraction \citep{Escala08,Genzel11}. With $M_{\mathrm{stars}} = (7.4\pm 1.2)\times 10^{10}~\mathrm{M}_{\odot}$ and $M_{\mathrm{molgas}} = (1.2\pm 0.2)\times 10^{10}~\mathrm{M}_{\odot}$, the A521-sys1 galaxy has $f_{\mathrm{gas}}$ smaller than 14 per cent as one also needs to account for the dark matter mass. The dynamical mass estimate, obtained under the disc-like potential hypothesis, does not show evidence for a large amount of dark matter ($\lesssim 5$ per cent). 
Moreover, given that $\sigma_0$ has not to be isotropic, $Q_{\mathrm{gas}}$ is likely close to 1 for the molecular gas disc, as expected since once disc fragmentation occurred, $Q_{\mathrm{gas}}$ should become $\geq 1$ to regulate the system.
Numerical simulations indeed show that the UV-bright stellar clumps, ubiquitously observed in galaxies at cosmic noon \citep[e.g.][]{Guo15,Dessauges17,Cava18,Zanella19}, form in gas clouds produced from the fragmentation process of high-redshift gas-rich, turbulent discs \citep[e.g.][]{Dekel09,Tamburello15,Mandelker17,Renaud21}. Feedback and molecular gas fraction play a key role here, since to undergo the phase of violent disc instability the feedback has not to be too strong \citep{Mayer16} and $f_{\mathrm{gas}}$ has to be at least $40-50$ per cent for fragmentation to happen and the formation of long-lived clouds \citep{Tamburello15,Fensch21}.

By combining the {\it ALMA} high angular resolution imaging capability with gravitational lensing, our study of the A521-sys1 strongly-lensed galaxy led to the second observational discovery of GMCs in a proto-Milky Way galaxy at $z\simeq 1$, after the Cosmic Snake galaxy \citep{Dessauges19}. Fourteen GMCs were identified, with 11 found to be gravitationally bound structures in virial equilibrium. They provide additional clues that GMCs hosted in typical high-redshift main-sequence galaxies are different from local GMCs. They have, on average, 100 times higher $M_{\mathrm{molgas}}$, 10 times higher $\Sigma_{\mathrm{molgas}}$, almost 100 times stronger internal kinetic pressures, and 10 times enhanced supersonic turbulence, resulting in being located above the Larson scaling relations for local GMCs \citep{Larson81,Bolatto08}. Only GMCs hosted in nearby starbursting and merger galaxies \citep[e.g.][]{Wei12,Leroy15,Rosolowsky21}, sometimes considered as the high-redshift galaxy analogues, show comparable physical properties. These findings support that, now and at other look-back times, GMCs form with properties that adjust to the ambient ISM conditions (pressure, turbulence, density, etc.) prevalent in the host galaxy.

This picture is further validated by the variations in the physical properties we observe among local GMCs \citep{Hughes13,Sun18,Sun20}, as well as among high-redshift GMCs hosted in different galaxies. Indeed, we may see that the 7 times weaker hydrostatic pressure that we estimated at the disc midplane of the A521-sys1 galaxy, in comparison to the one of the Cosmic Snake galaxy, seems to impact in cascade all the properties of the A521-sys1 GMCs with respect to those of the Cosmic Snake GMCs: it first implies a weaker GMC internal kinetic pressure, which yields on the one hand i)~a lower gas mass surface density (Fig.~\ref{fig:pressure}) and on the other hand ii)~a less extreme supersonic turbulence given the lower internal velocity dispersion (equation~(\ref{eq:Mach-number})). Because of i) GMCs results in being less shielded from the ambient photodissociating radiation and therefore we measured a smaller CO-to-H$_2$ conversion factor from the virialized clouds (Sect.~\ref{sect:pressure}), and because of ii) GMCs are found to be slightly less efficient in forming stars (Sect.~\ref{sect:SFE}).

In summary, the GMCs, detected in the A521-sys1 galaxy and the Cosmic Snake galaxy, have masses high enough to support the in-situ formation of the massive stellar clumps (star cluster complexes/associations) we see in comparable numbers in the {\it HST} rest-frame UV to optical images of these two $z\simeq 1$ galaxies \citep{Messa22,Cava18}, as well as in up to $\sim 60$ per cent of galaxies at $z\sim 1- 2$ \citep{Guo15}. Their similar molecular gas and stellar mass distributions suggest a relatively high star-formation efficiency of $\sim 30$ per cent in these $z\simeq 1$ GMCs. We estimated the star-formation efficiency per free-fall time to be $\sim 11$ per cent in the A521-sys1 galaxy. To confirm the possible trend of an enhanced star-formation efficiency in GMCs of high-redshift galaxies, high-resolution H$\alpha$ observations, tracing the very recent ($\lesssim 10~\mathrm{Myr}$) star formation, are needed, as motivated by the studies of GMC star-formation time-scales and efficiencies led in nearby galaxies \citep[e.g.][]{Kruijssen19,Chevance20}, based on the statistical framework developed by \citet{Kruijssen14} and \citet{Kruijssen18}, built on the spatial correlation/decorrelation of GMCs and young stars on small scales.
Planned {\it JWST} Cycle~1 H$\alpha$ integral-field NIRSpec observations of the Cosmic Snake galaxy will help to shed light on this important physical parameter so far usually fixed to $\epsilon_{\mathrm{ff}}\sim 1$ per cent as measured in nearby galaxies \citep{Krumholz12}.


\section*{Acknowledgements}

The data analysed in this paper were obtained with the {\it ALMA} Observatory, under the program 2016.1.00643.S. {\it ALMA} is a partnership of ESO (representing its member states), NSF (USA) and NINS (Japan), together with NRC (Canada), MOST and ASIAA (Taiwan), and KASI (Republic of Korea), in cooperation with the Republic of Chile. The Joint {\it ALMA} Observatory is operated by ESO, AUI/NRAO and NAOJ.
M.M. and A.A. acknowledge the support of the Swedish Research Council, Vetenskapsr\r{a}det project grants 2019-00502 and 2021-05559, respectively. 


\section*{Data Availability}


The data underlying this article will be shared on reasonable request to the corresponding author.



\appendix

\section{Molecular cloud identification in channel maps}

\begin{figure*}
\includegraphics[width=0.33\textwidth,clip]{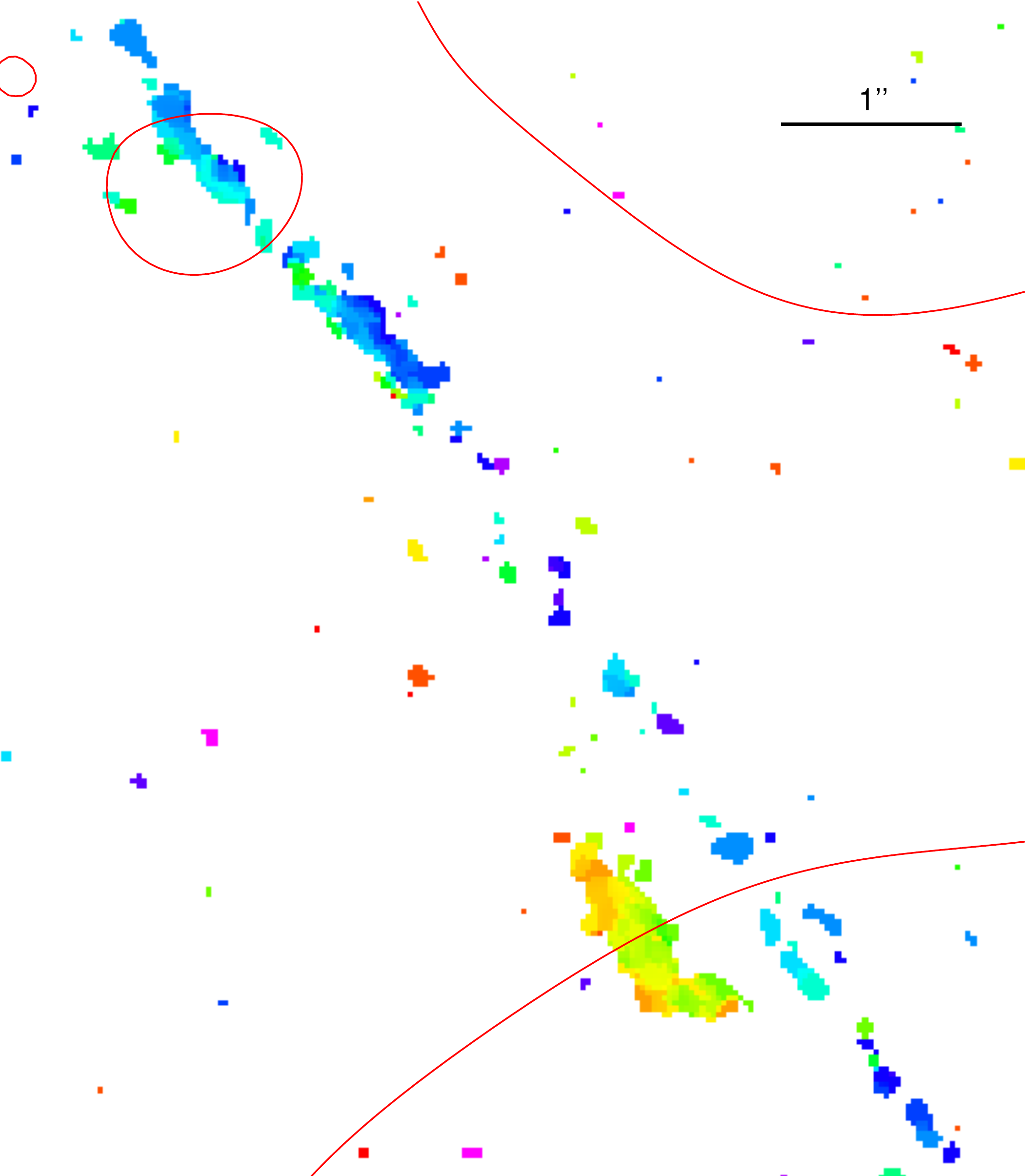}
\includegraphics[width=0.33\textwidth,clip]{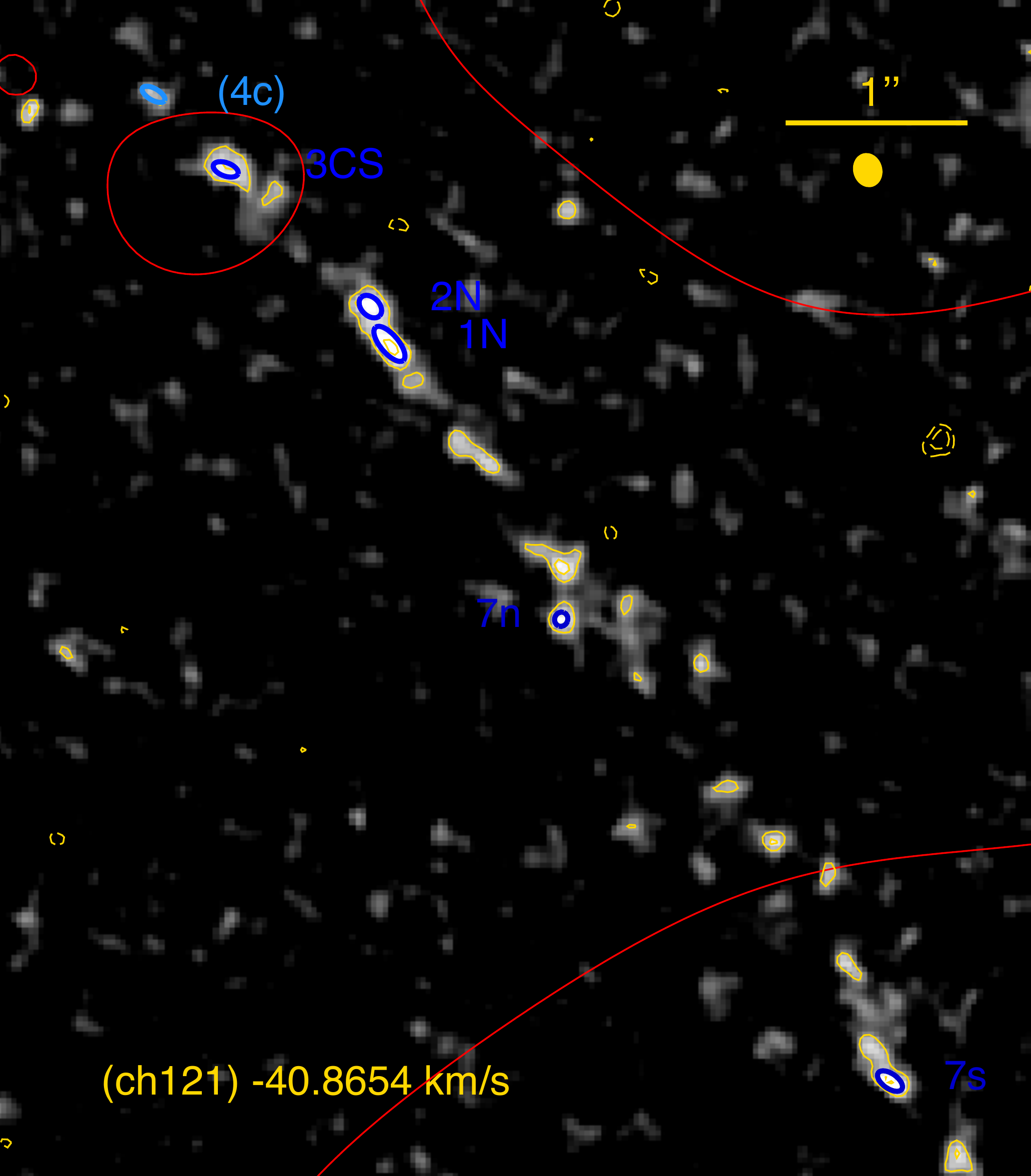}
\includegraphics[width=0.33\textwidth,clip]{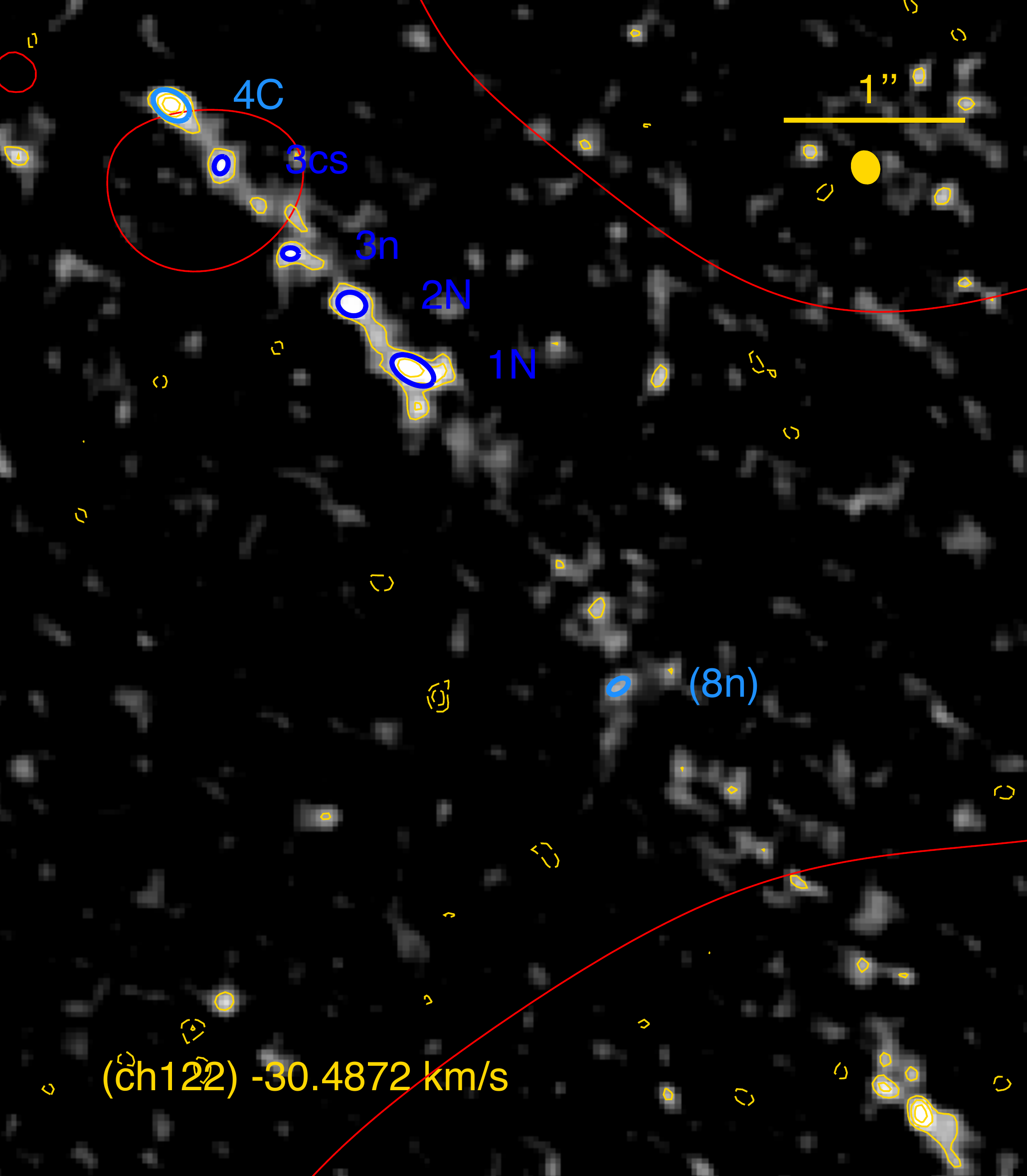}

\includegraphics[width=0.33\textwidth,clip]{figure-moment1_10-155kms.pdf}
\includegraphics[width=0.33\textwidth,clip]{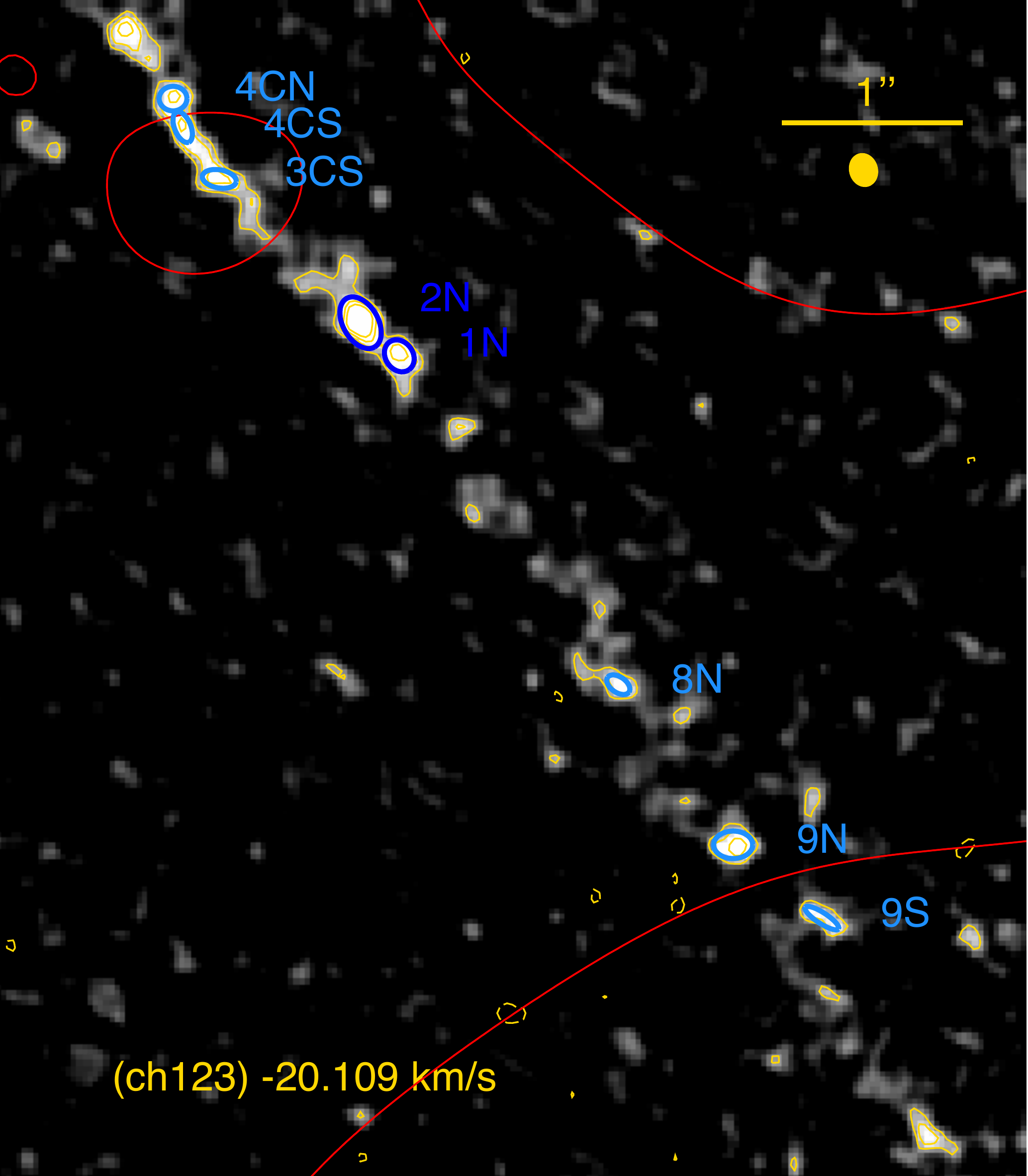}
\includegraphics[width=0.33\textwidth,clip]{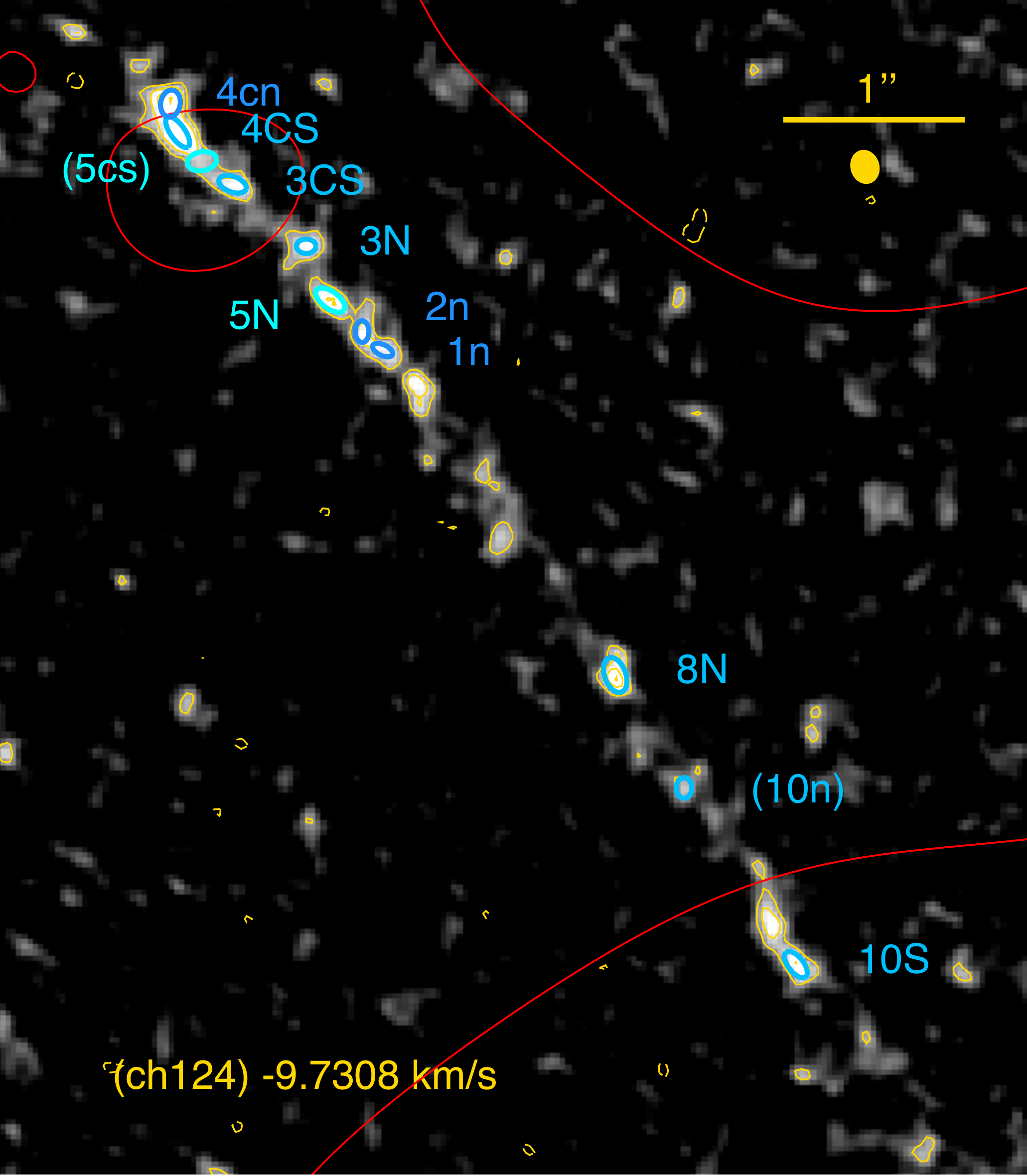}

\includegraphics[width=0.33\textwidth,clip]{figure-moment1_10-155kms.pdf}
\includegraphics[width=0.33\textwidth,clip]{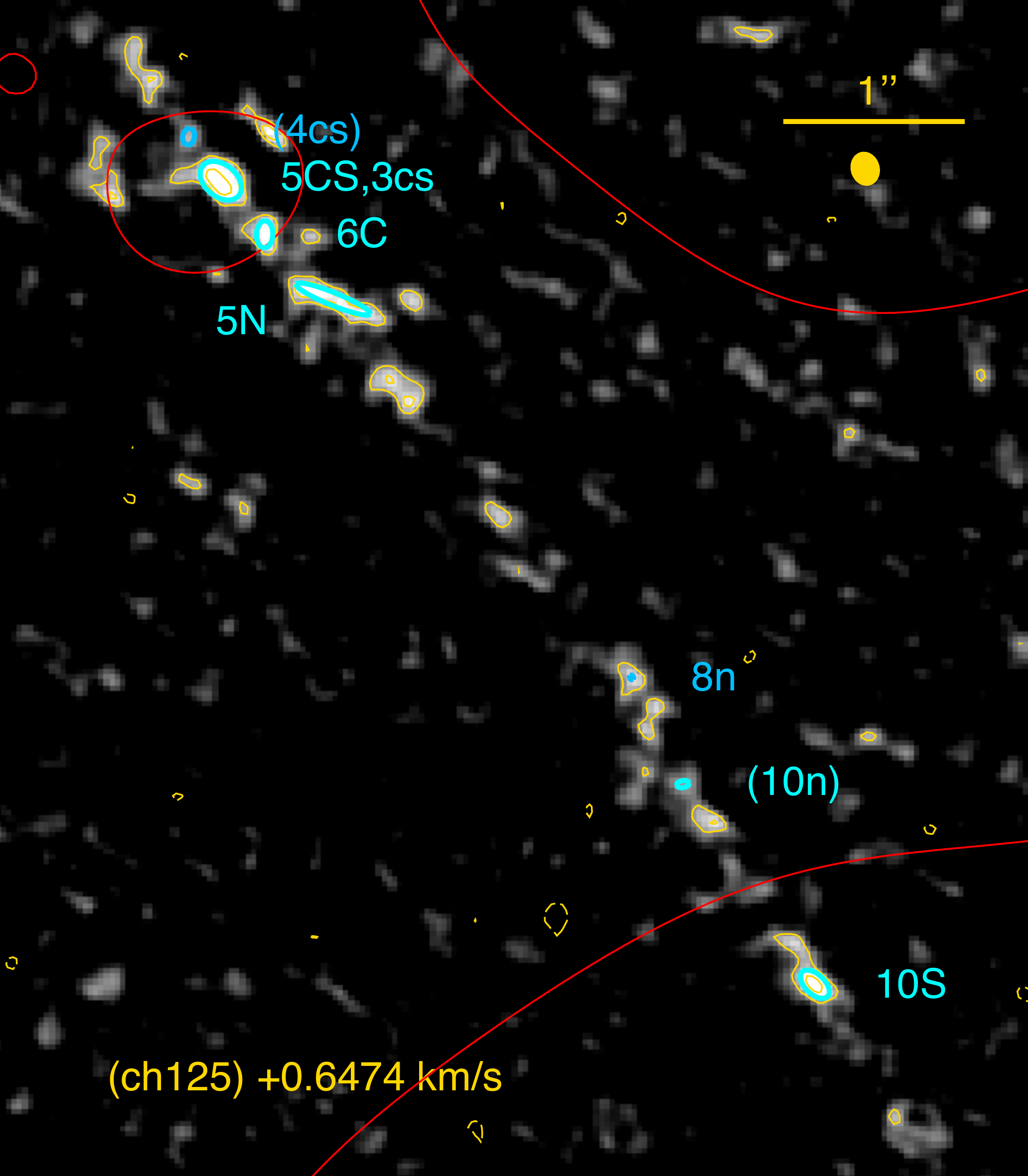}
\includegraphics[width=0.33\textwidth,clip]{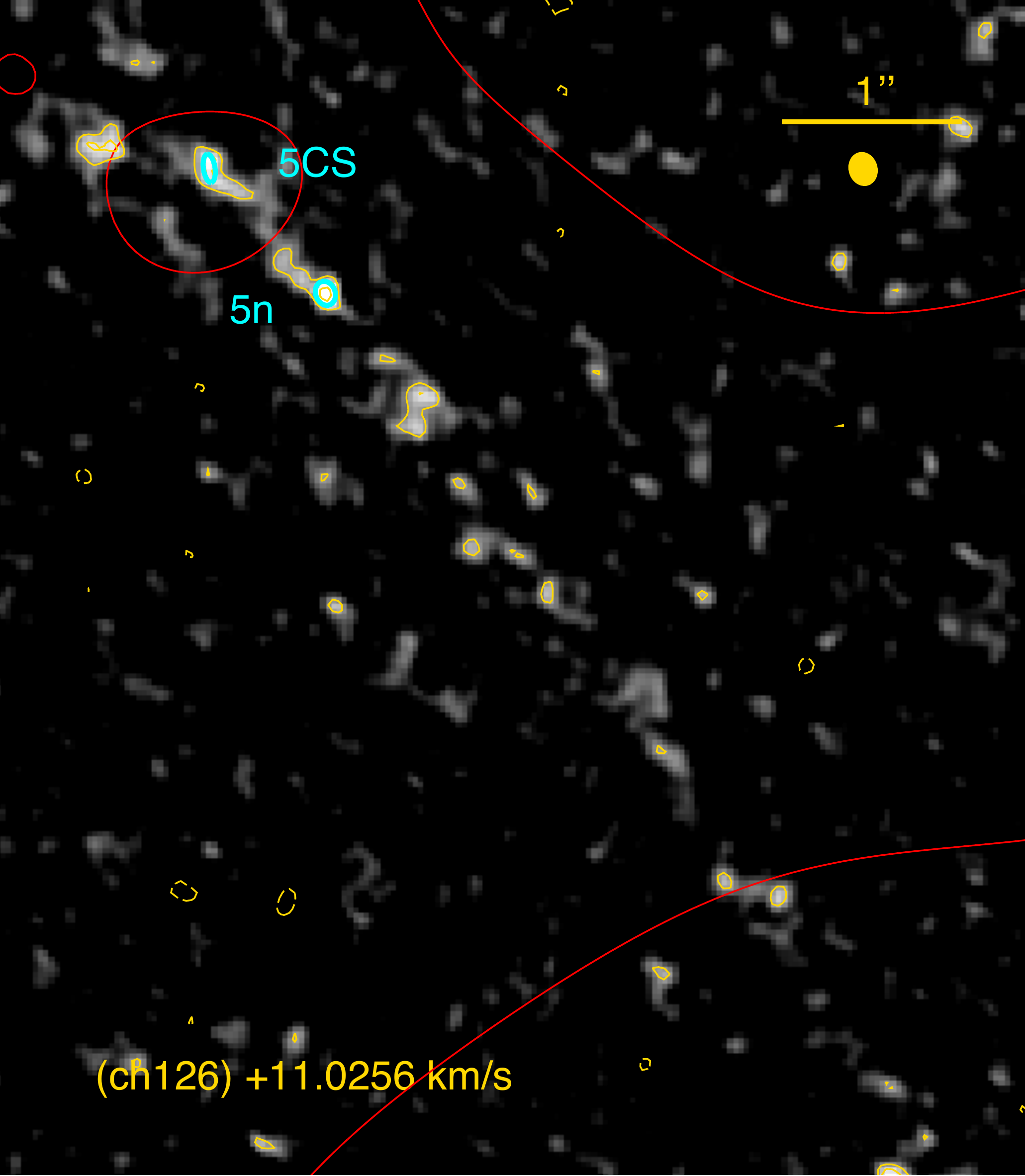}
\caption{In each row, the left panel shows the velocity first-moment map, 
with the magenta/violet colors corresponding to $-70~\mathrm{km~s^{-1}}$ and the red colors to $+90~\mathrm{km~s^{-1}}$. The critical line of our lens model at the redshift of the A521-sys1 galaxy, $z=1.043$, is represented by the red solid line. The middle and right panels show the channel intensity maps of the CO(4--3) line emission at the native spectral resolution of $10.3782~\mathrm{km~s^{-1}}$. The overlaid gold contours start at $\pm 3\sigma$, in steps of $1\sigma$ ($\mathrm{RMS = 0.0027~Jy~beam^{-1}~km~s^{-1}}$); dashed gold contours for negative values. The ellipses, color-coded following the velocity map at their location, are the best-fits of the extracted $4\sigma$ intensity contours (sometimes the $3\sigma$ intensity contours (for labels in parenthesis) if clearly identified, or higher intensity contours if the $4\sigma$ intensity contours are blended) of the 24 counter-images of the 14 GMCs. The $3\sigma$ intensity contours with equivalent circularised radii bigger than the equivalent circularised radius of the beam, are marked with capital letters (if smaller they are marked with small letters).
The {\it ALMA} synthesized beam with a size of $0.19''\times 0.16''$ at the position angle of $-74^{\circ}$ is shown by the gold filled ellipse. We give a reference scale of 1~arcsec.}
\label{fig:Appendix}
\end{figure*}

In Sect.~\ref{sect:GMC-search} we describe how we performed the search for the GMCs in the {\it ALMA} 12-channel intensity maps extracted over the CO(4--3) line emission. In total 24 counter-images of 14 GMCs were identified, with most GMCs having two to three multiple images. Their respective CO(4--3) emission encompasses between one and five adjacent $10.3782~\mathrm{km~s^{-1}}$ channels, as shown in Fig.~\ref{fig:Appendix}. Similarly to the definition applied for the GMCs hosted in the Cosmic Snake galaxy \citep{Dessauges19}, the detected $4\sigma$ CO(4--3) emission peaks per channel were considered as belonging to images of distinct clouds when their $4\sigma$ contours were not spatially overlapping in the same channel map, or when their $4\sigma$ contours were co-spatial, but not in adjacent channel maps (that is, when separated by at least one channel). This definition of clouds, although at first sight appearing as arbitrary, is a possible way of defining clouds. We consider that the essential point is to describe our method clearly to ensure that our results are reproducible, but, of course, a different cloud definition can be adopted using the same {\it ALMA} data. Globally, we acknowledge that our definition of clouds depends on, and is limited by, our spatial resolution and our beam size. We labelled the identified multiple images with `N', `S', and `C' letters, they denote the respective emission peaks found in the northern (N) mirrored image and the southern (S) mirrored image of the arc, and the central (C) north (CS) and south (CS) images around the closed critical line in the northern part of the arc. The numbers enumerate the 14 distinct GMCs. 

In Fig.~\ref{fig:Appendix-spectra} we show the individual integrated CO(4--3) emission line spectra extracted for the 14 GMCs. We plot the spectra obtained for the less blended multiple images in channel maps of each GMC,
and that was adopted to estimate the GMC physical properties listed in Table~\ref{tab:GMCs}.

\begin{figure*}
\includegraphics[width=0.33\textwidth,clip]{figure-moment1_10-155kms.pdf}
\includegraphics[width=0.33\textwidth,clip]{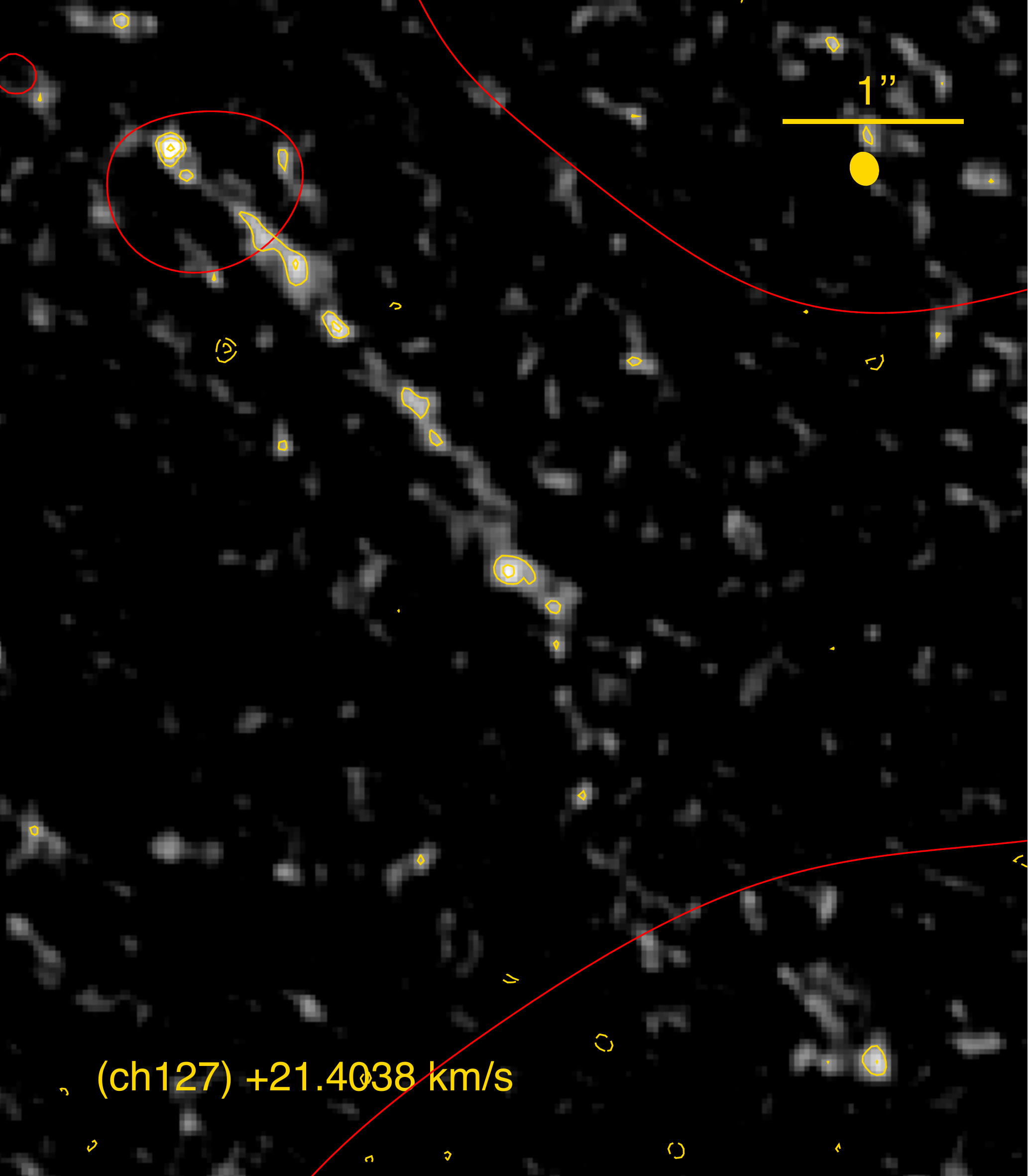}
\includegraphics[width=0.33\textwidth,clip]{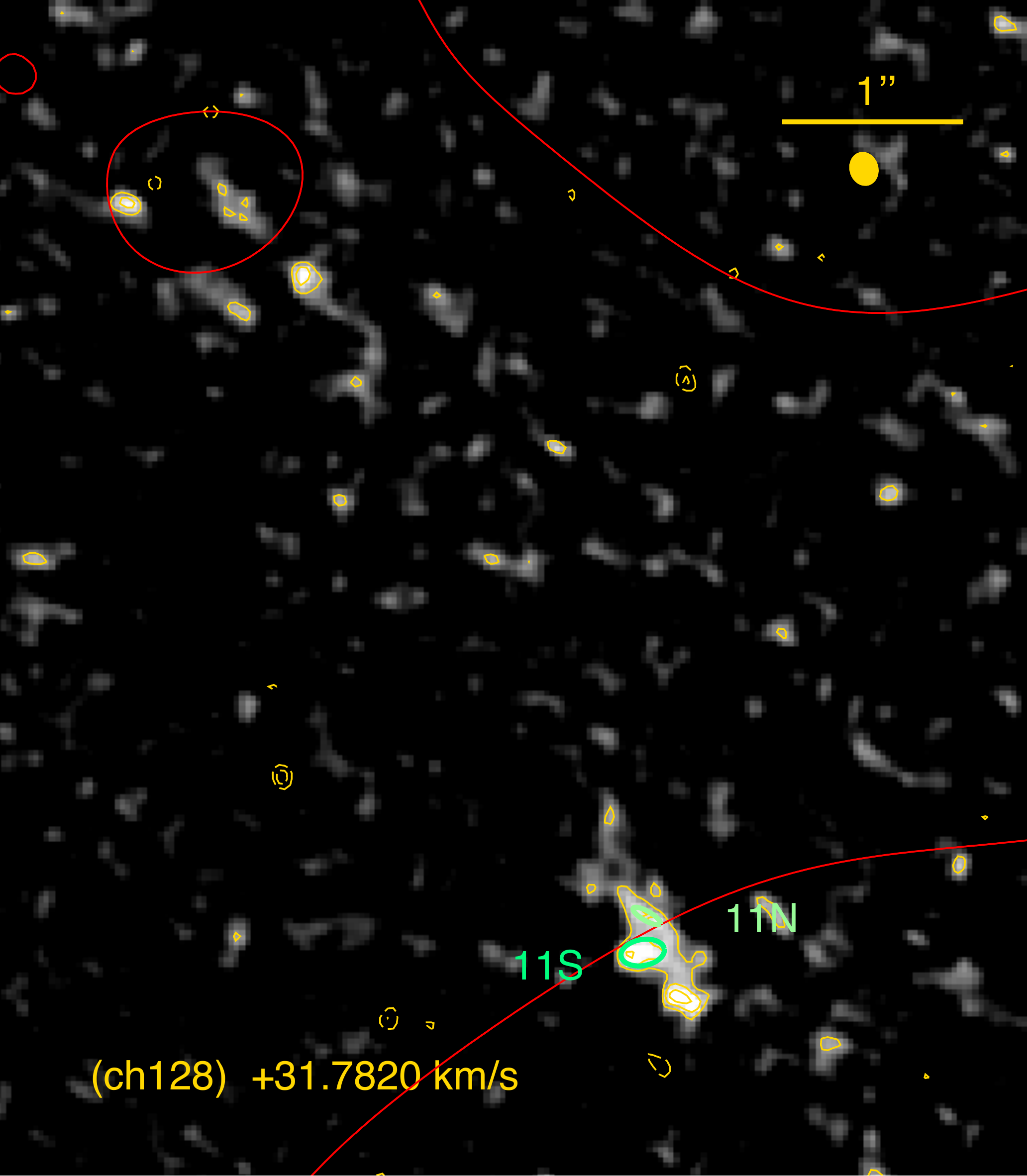}

\includegraphics[width=0.33\textwidth,clip]{figure-moment1_10-155kms.pdf}
\includegraphics[width=0.33\textwidth,clip]{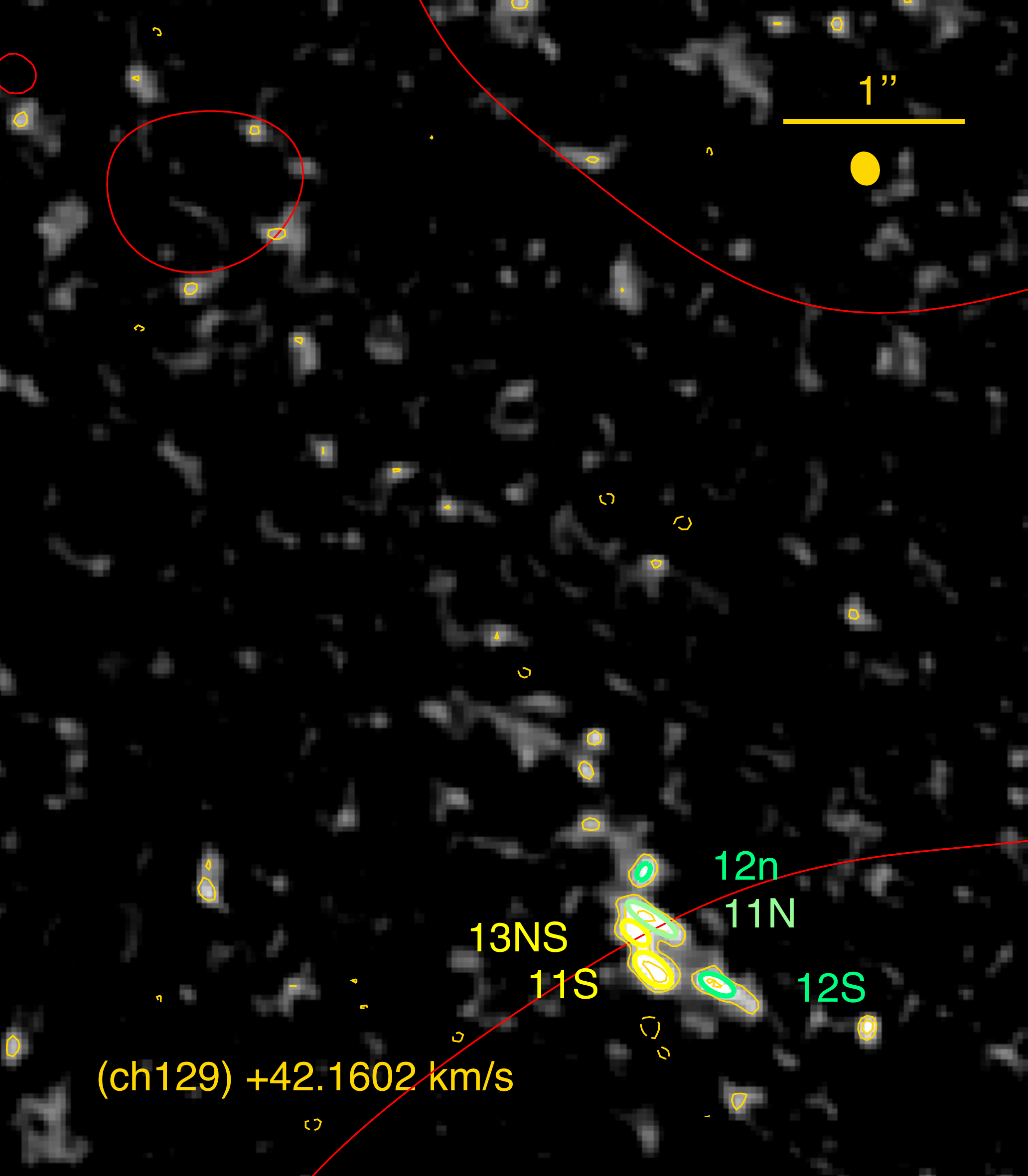}
\includegraphics[width=0.33\textwidth,clip]{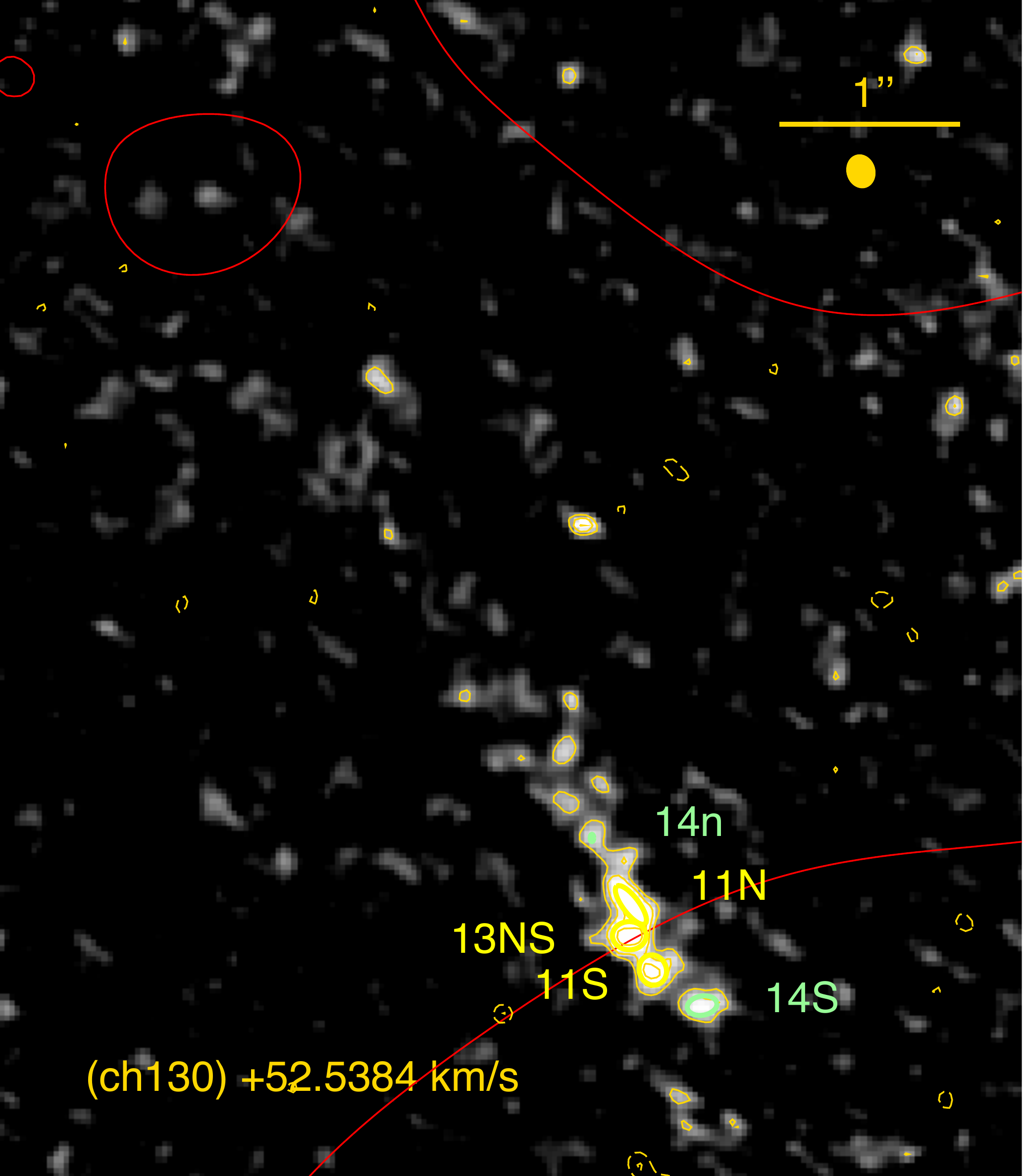}

\includegraphics[width=0.33\textwidth,clip]{figure-moment1_10-155kms.pdf}
\includegraphics[width=0.33\textwidth,clip]{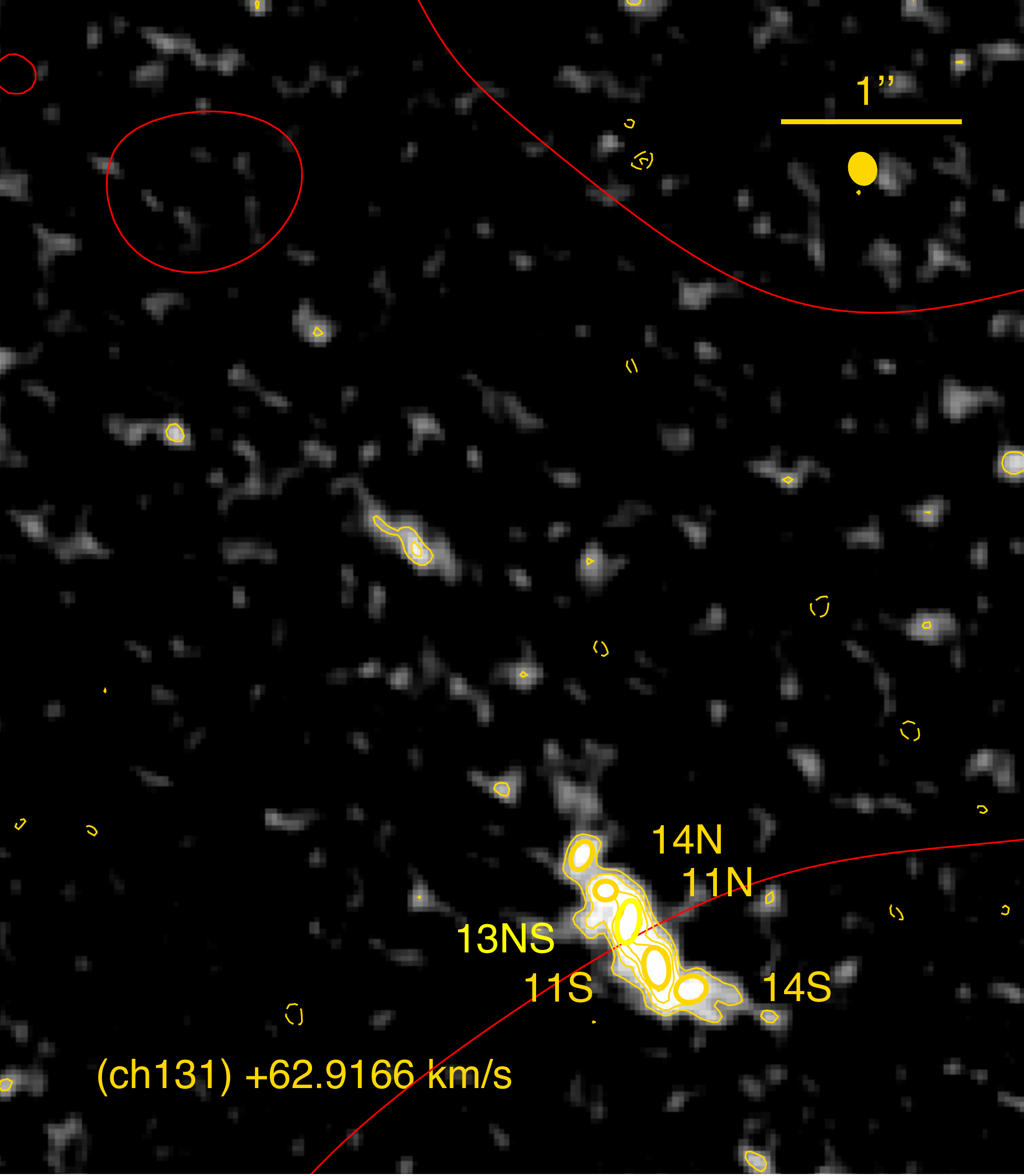}
\includegraphics[width=0.33\textwidth,clip]{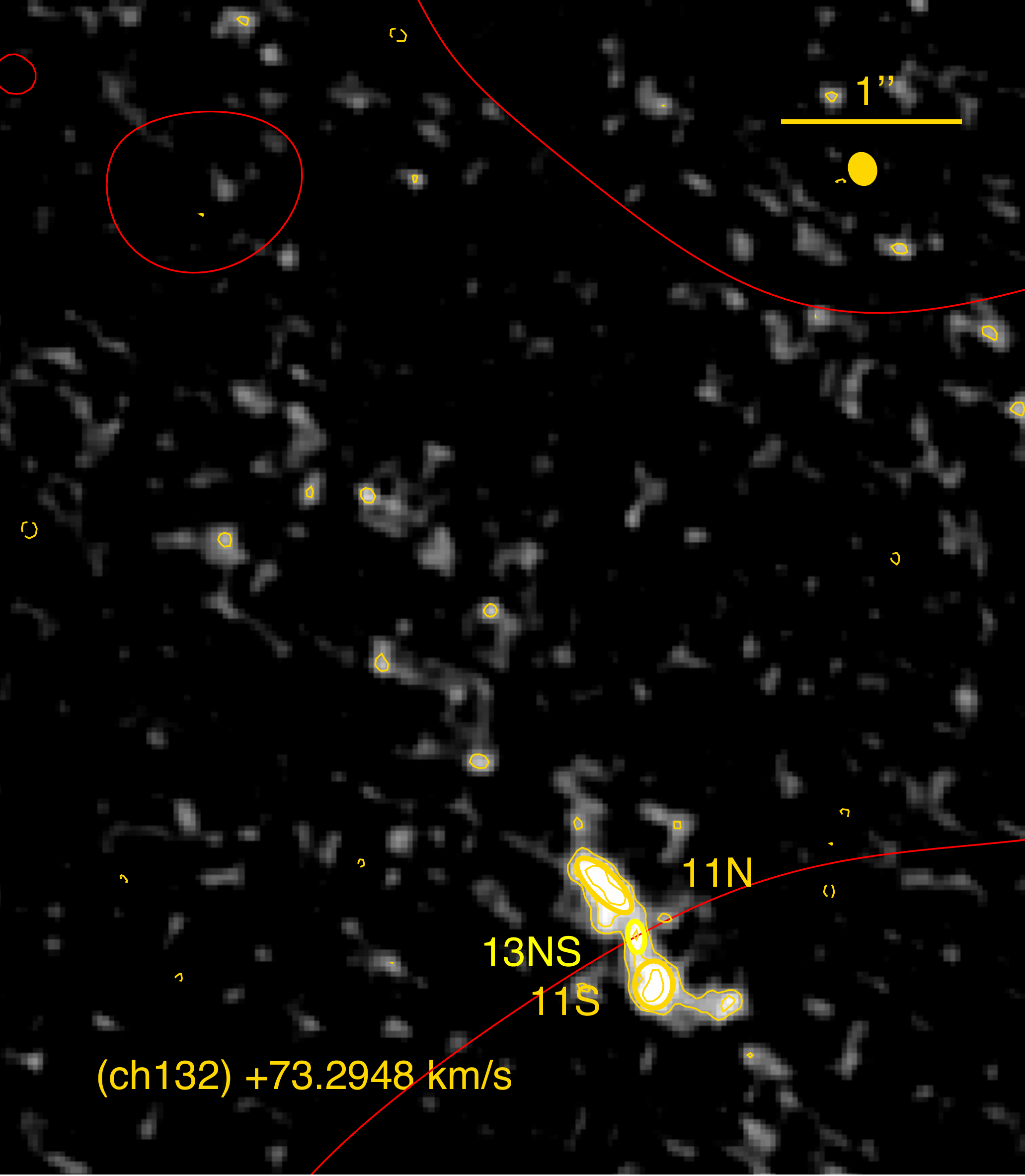}
\smallskip
\centerline{{\bf Figure A1} -- {\it continued}}
\smallskip
\end{figure*}
\begin{figure*}
\includegraphics[width=0.8\textwidth,clip]{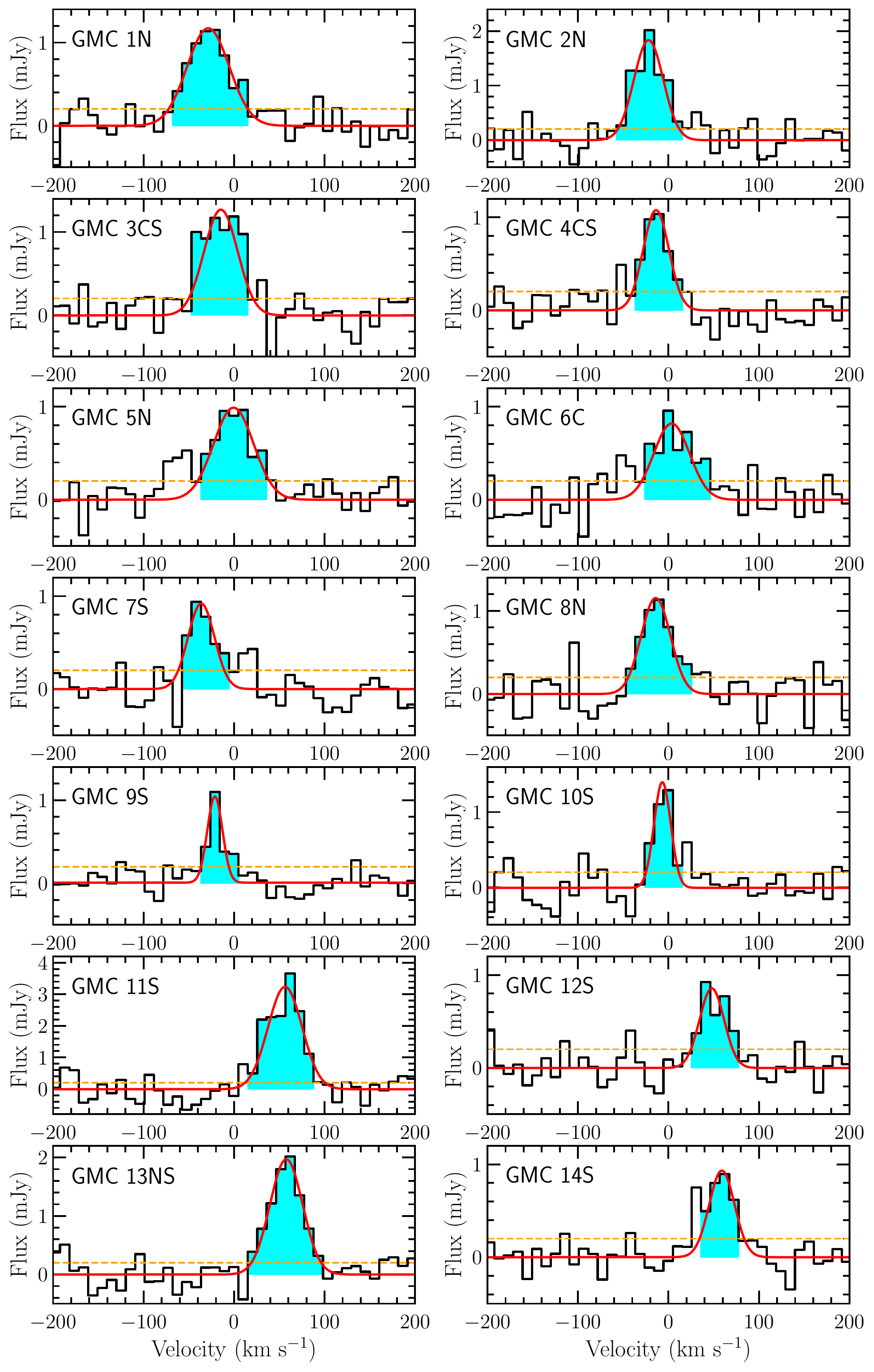}
\caption{{\it ALMA} integrated CO(4--3) emission line spectra (plotted at the spectral resolution of $10.3782~\mathrm{km~s^{-1}}$) of the 14 GMCs detected in the A521-sys1 galaxy, extracted from the less blended multiple images in channel maps (Fig.~\ref{fig:Appendix}) of each GMC. The cyan shaded regions represent channels over which the CO(4--3) emission is detected above the RMS noise of $0.2~\mathrm{mJy~beam^{-1}}$ (orange dashed line).
The best Gaussian fits of the CO(4--3) line profiles are shown by the red solid lines. The zero velocity is fixed at $z=1.043425$.}
\label{fig:Appendix-spectra}
\end{figure*}

\section{CASA visibility simulations}

To validate the existence of the detected GMCs (see Sect.~\ref{sect:GMC-search}) against artificial clumpy features produced by lensing and interferometric effects out of a smooth exponential disc of CO(4--3) emission, we followed the methodology of \citet{Hodge16,Hodge19}, \citet{Gullberg18}, and \citet{Ivison20} who evidenced the presence of artificial clumpy features in lensed interferometric data. We performed CASA visibility simulations with the CASA \texttt{simobserve} routine by setting up the frequency, the bandwidth, the ALMA configuration, the precipitable water vapour, and the exposure time to the real ALMA observations of A521-sys1. For the CO(4--3) line imaging, we used the CASA \texttt{simanalyze} routine and we applied the same pixel size and weighting scheme as for real visibilities using Briggs and the robust factor of 0.5. The generated RMS noise per beam in the simulated channel intensity maps is comparable within about 10 per cent to the RMS noise measured in the real channel intensity maps for the same synthesized beam size. We considered two input sky models. 

\begin{figure*}
\includegraphics[width=0.245\textwidth,clip]{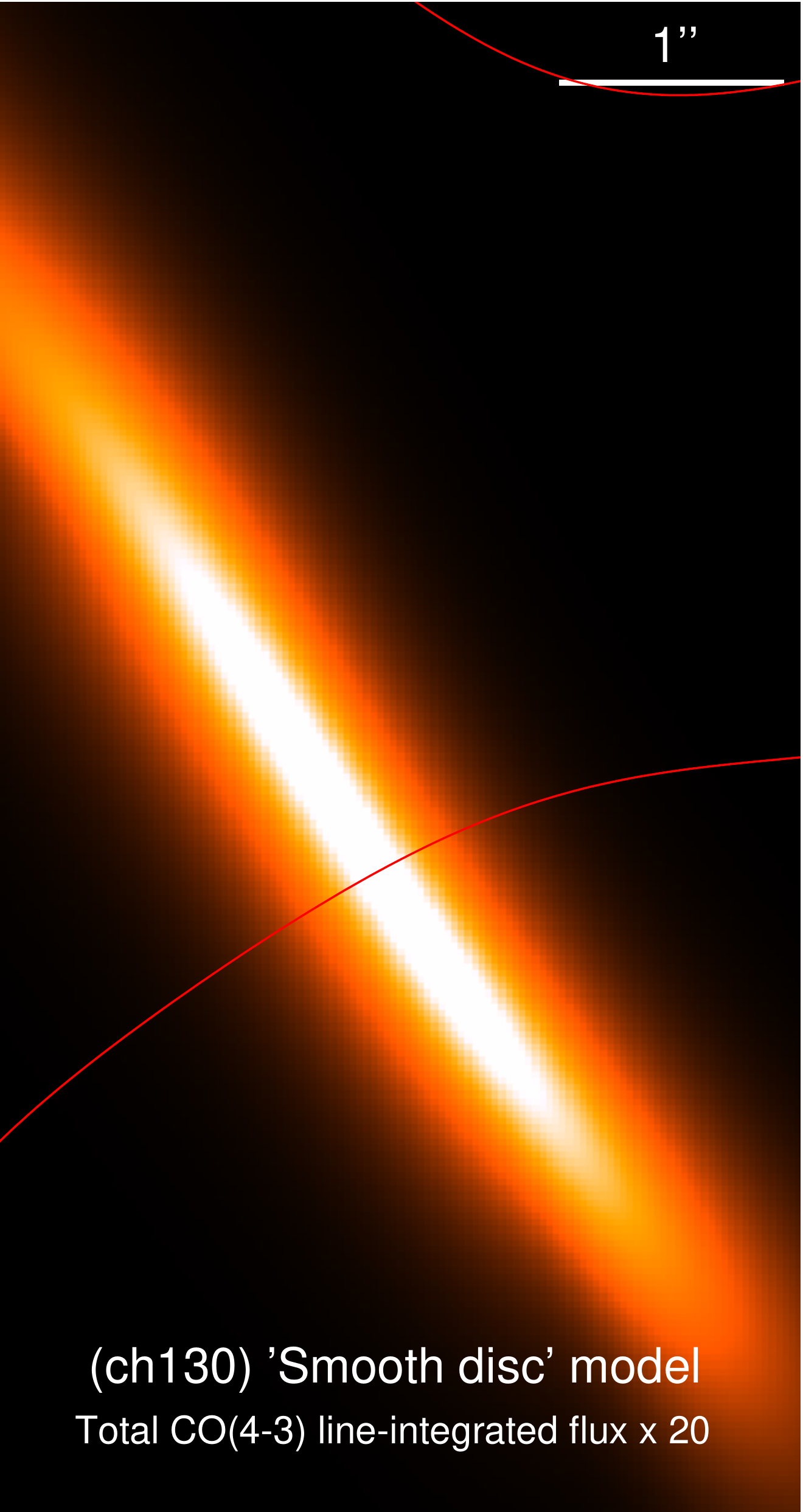}
\includegraphics[width=0.245\textwidth,clip]{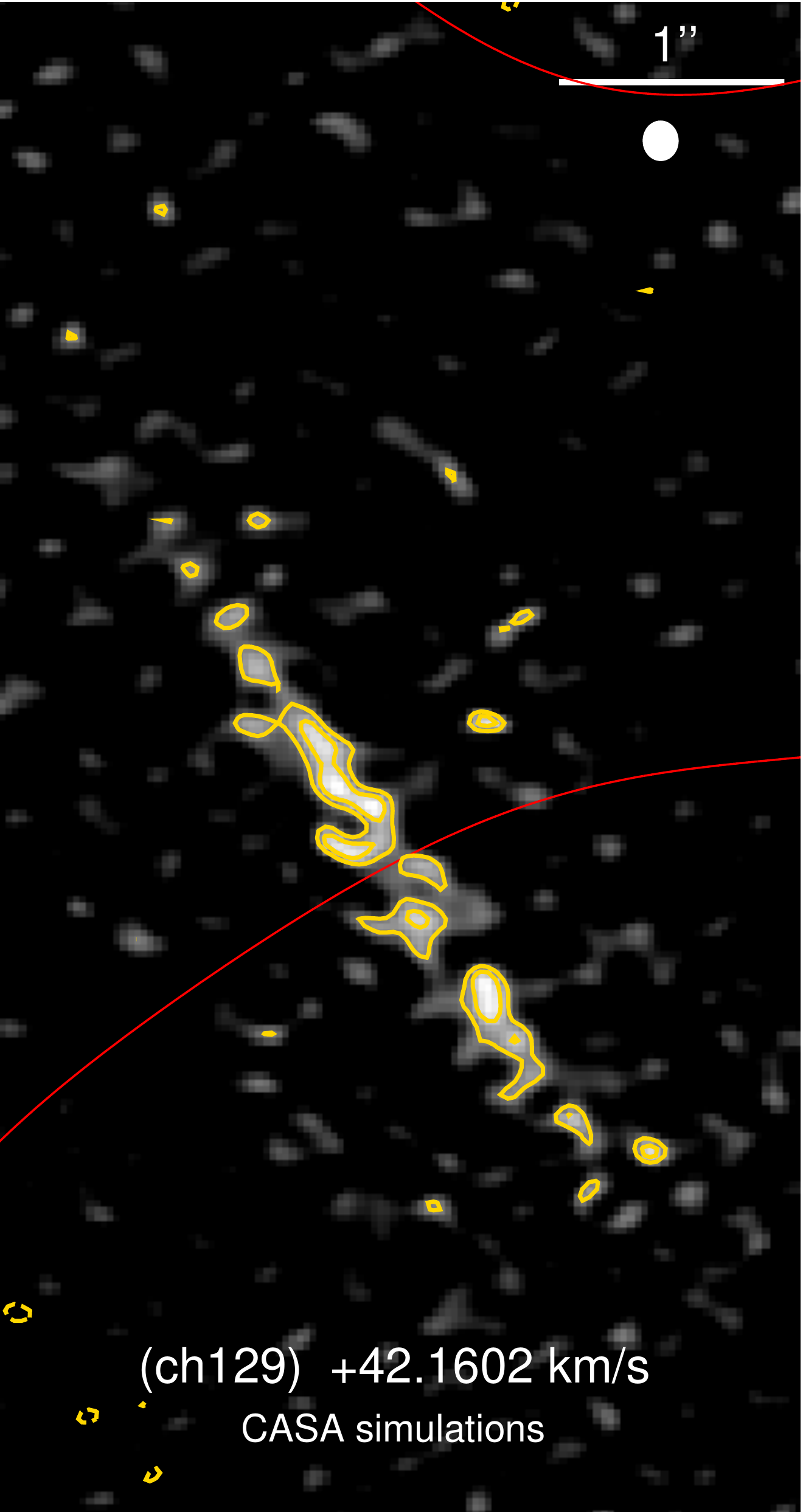}
\includegraphics[width=0.245\textwidth,clip]{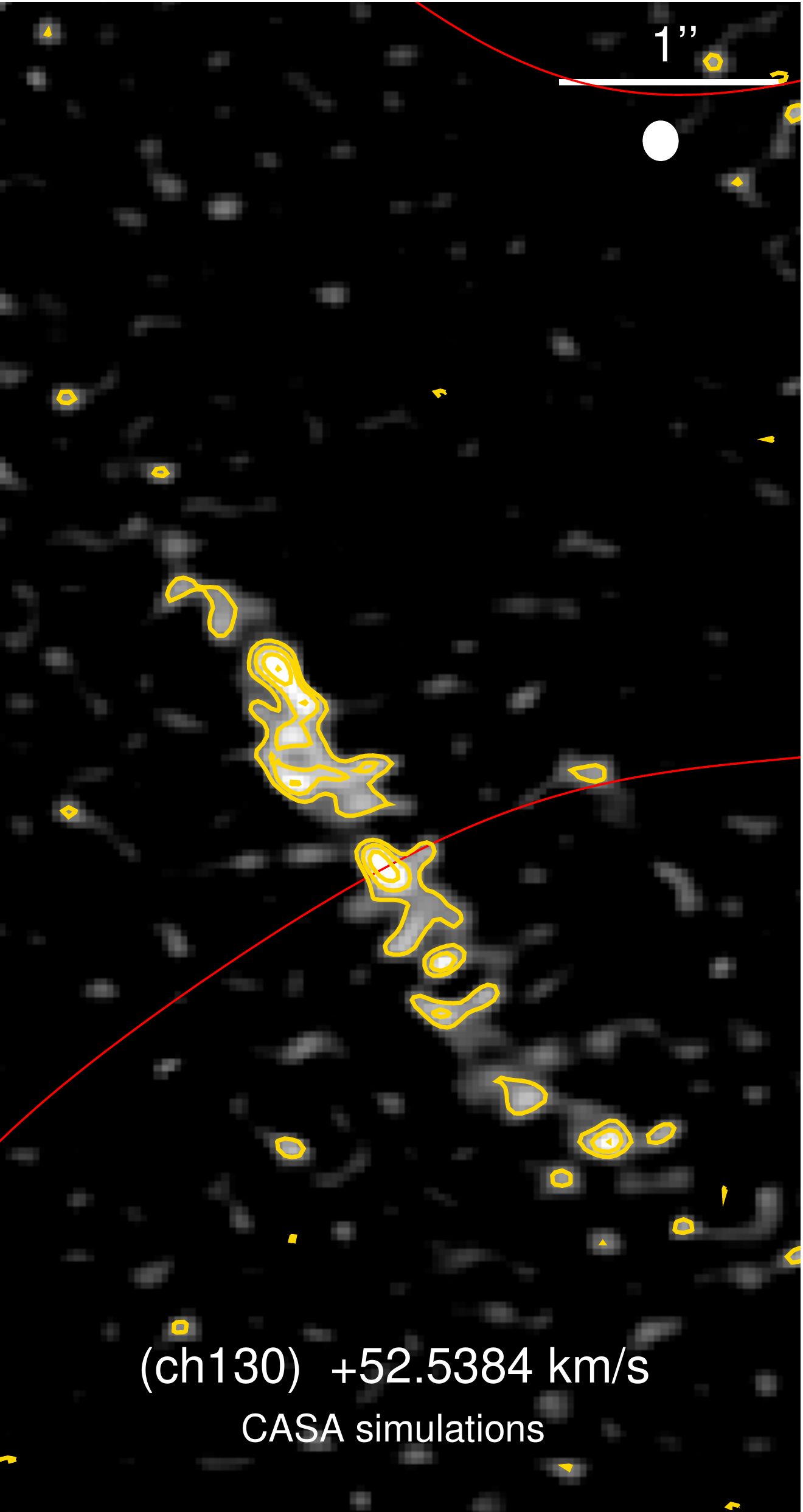}
\includegraphics[width=0.245\textwidth,clip]{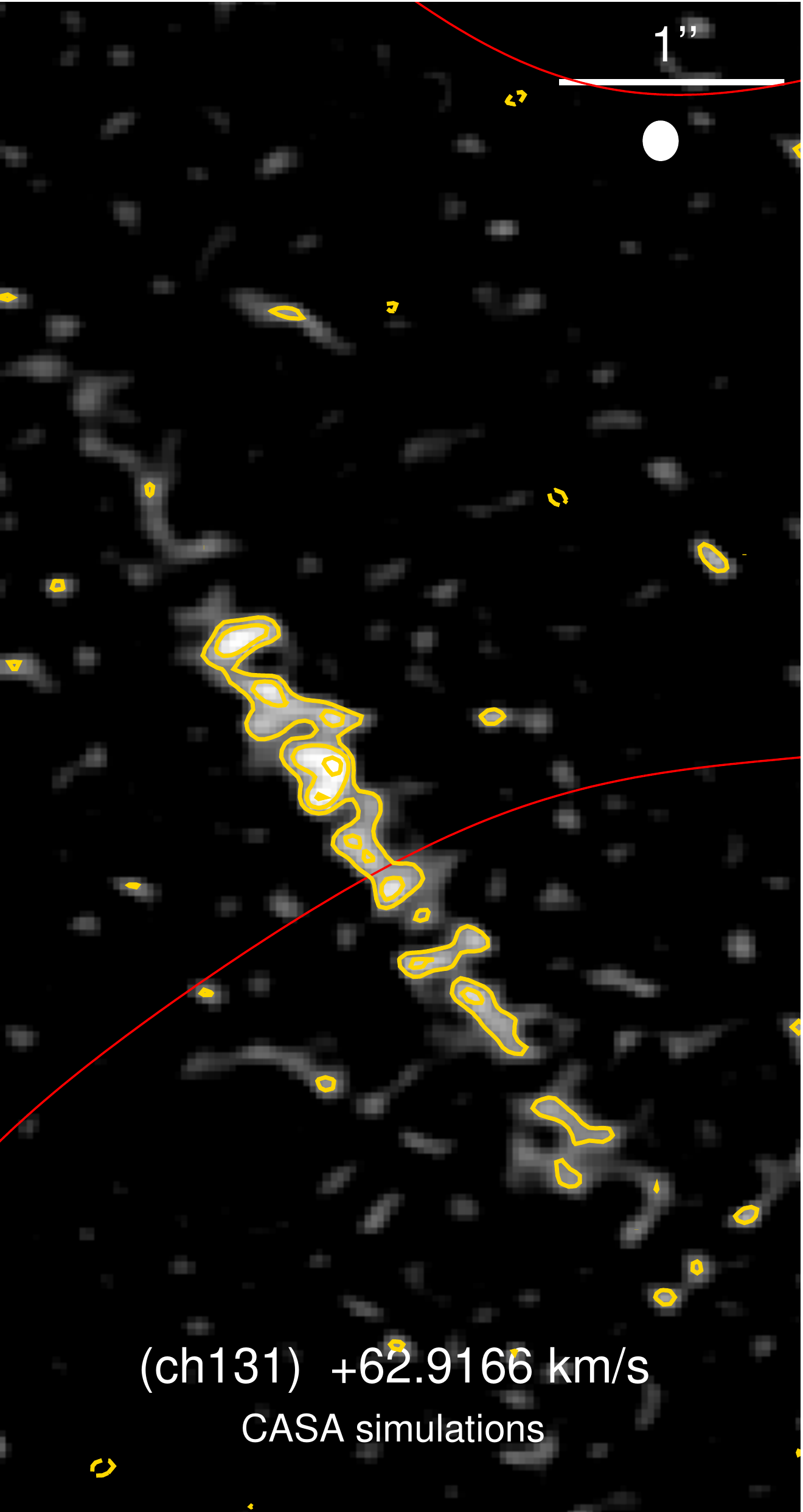}
\caption{{\it ALMA} visibility simulations of a lensed smooth exponential disc with the S\'ersic index of 1, characterised by the total CO(4--3) line-integrated flux multiplied by 20, the effective radius, and the rotation velocity measured for the A521-sys1 galaxy. The left panel shows an example of the channel intensity map of $10.3782~\rm km~s^{-1}$ of the central bright region of the smooth disc model located close to the critical line. 
The other panels show the corresponding CASA simulated intensity maps in three consecutive channels with the RMS noise comparable to the real {\it ALMA} observations of A521-sys1. Only with the flux boost by a factor of 20, the CO(4--3) emission of the smooth disc model is detected at $4-6\sigma$ above the RMS noise in the simulated channel maps and appears structured in clumpy features as shown by the overlaid gold contours starting at $\pm 3\sigma$, in steps of $1\sigma$ ($\rm RMS = 0.003~Jy~beam^{-1}~km~s^{-1}$); dashed gold contours for negative values. The clumpy features mimic the GMCs we identified in channel maps (see Fig.~\ref{fig:Appendix}), but remain fundamentally different because they are found at random locations from channel to channel and they change for different noise realisations. 
The critical line of our lens model at the redshift of the A521-sys1 galaxy, $z=1.043$, is shown by the red solid line. The $0.18''\times 0.15''$ synthesized beam of the simulated data at the position angle of $-89^{\circ}$ is represented by the white filled ellipse. We give a reference scale of 1 arcsec.}
\label{fig:Appendix-simu_smoothdisk}
\end{figure*}

In the first sky model, we distributed the total lensing-corrected CO(4--3) line-integrated flux of $1.5\pm 0.2~\rm Jy~km~s^{-1}$ measured in the A521-sys1 galaxy in a smooth exponential disc with the S\'ersic index of 1 in the source plane, characterised by the rotation velocity and the effective radius determined for the A521-sys1 galaxy \citep{Girard19,Nagy22}. This smooth exponential disc was then lensed with the tailored lens model using Lenstool (see Sect.~\ref{sect:lens-model}) to obtain the corresponding CO(4--3) emission data cube in the image plane, for which we adopted a channel sampling of $10.3782~\rm km~s^{-1}$ as in the real data cube. The simulated channel intensity maps showed no CO(4--3) emission detection of the smooth disc model. Only after increasing the measured total CO(4--3) line-integrated flux by a factor of 20, the CO(4--3) emission of the smooth disc model was detected at $4-6\sigma$ above the RMS noise in the simulated channel maps; albeit only in the central bright regions of the disc located close to the critical line. The detected CO(4--3) emission is indeed structured in clumpy features as shown in Fig.~\ref{fig:Appendix-simu_smoothdisk} for three consecutive channels encompassing the central bright region of the smooth disc. However, we noticed that these clumpy features change for different noise realisations in our CASA visibility simulations; as a result, the corresponding CO emission peaks that mimic the GMC search undertaken per channel map, have random locations, and are thus never detected over two and more adjacent channels at the same location and do not show multiple images as predicted by the lens model, while the GMCs identified in the real channel maps do (see Sect.~\ref{sect:GMC-search} and Fig.~\ref{fig:Appendix}). Moreover, and most importantly, it is difficult to explain how we could have missed 20 times of the CO(4--3) line-integrated flux of A521-sys1, and this even more that, in addition to the {\it ALMA} observations, we have {\it IRAM} 30~m single dish observations that do not suffer of large-scale CO(4--3) emission filtering.

\begin{figure*}
\includegraphics[width=0.245\textwidth,clip]{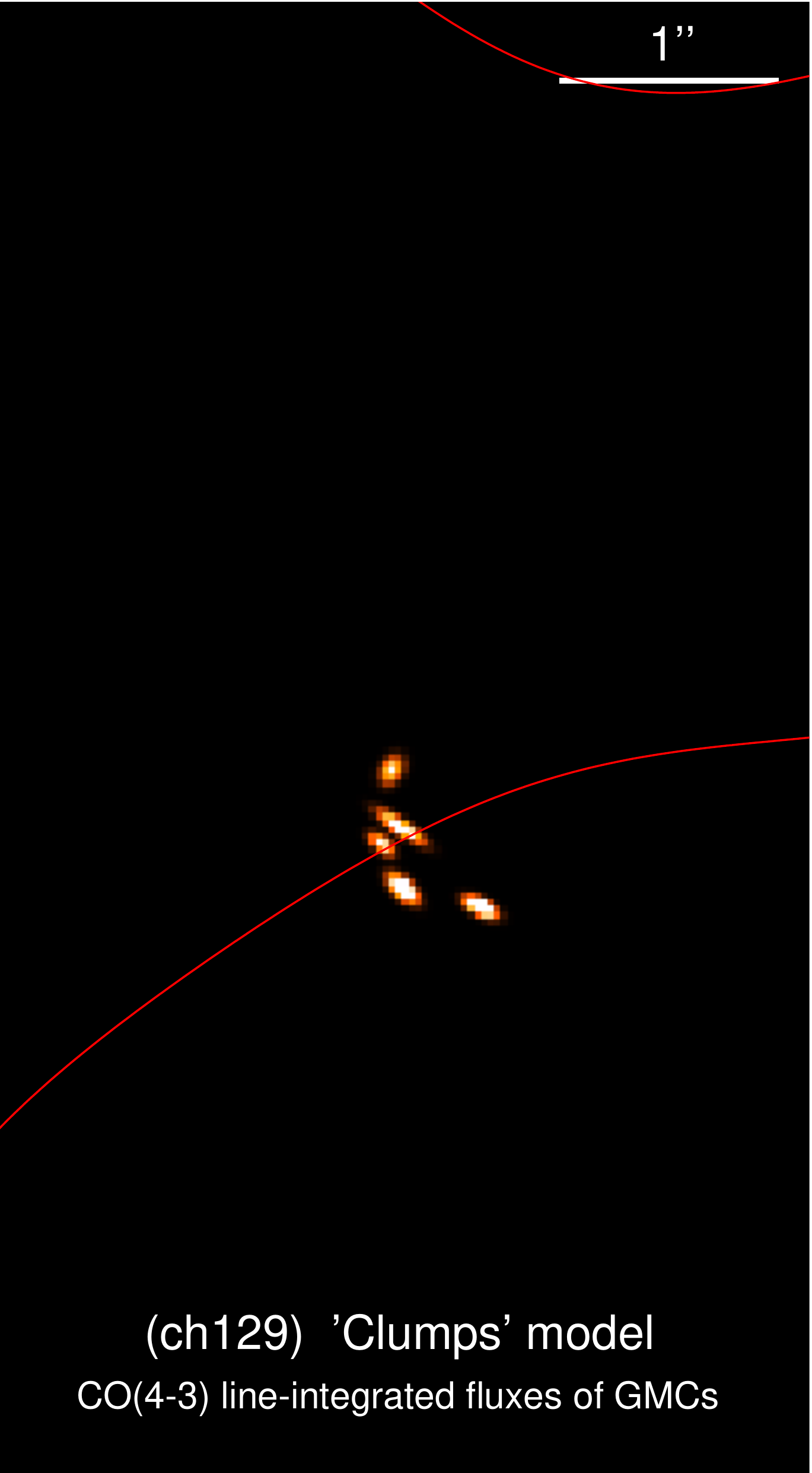}
\includegraphics[width=0.245\textwidth,clip]{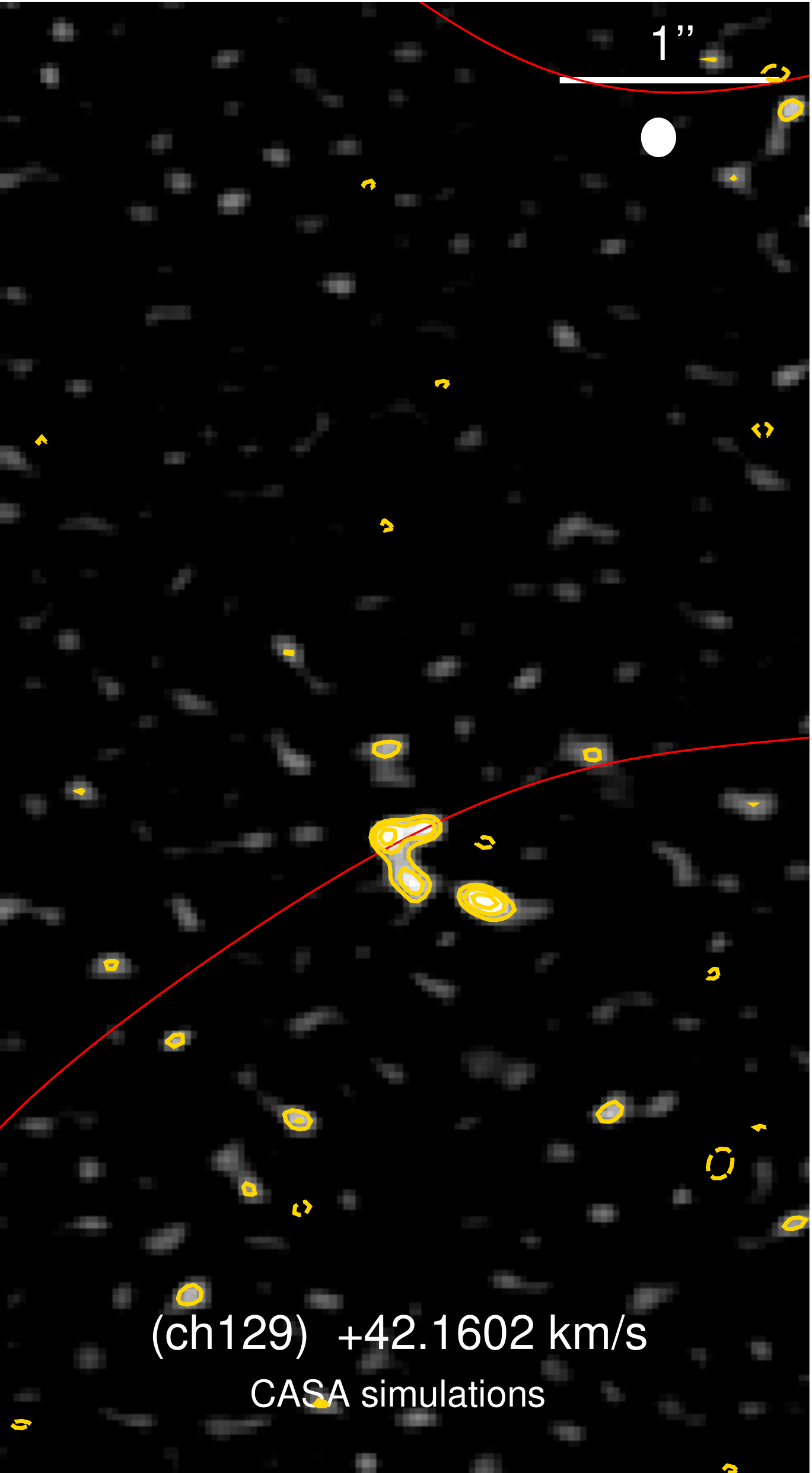}
\includegraphics[width=0.245\textwidth,clip]{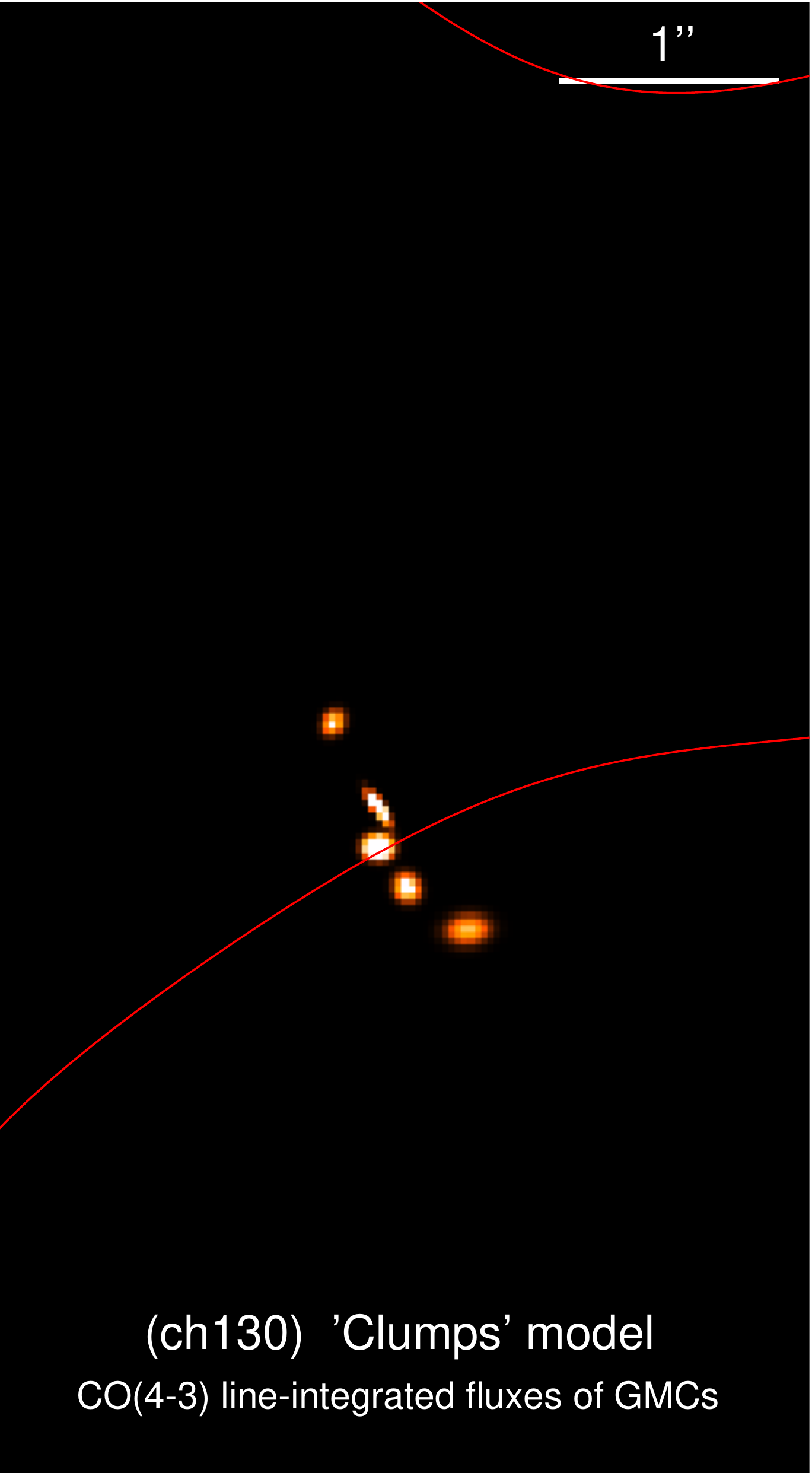}
\includegraphics[width=0.245\textwidth,clip]{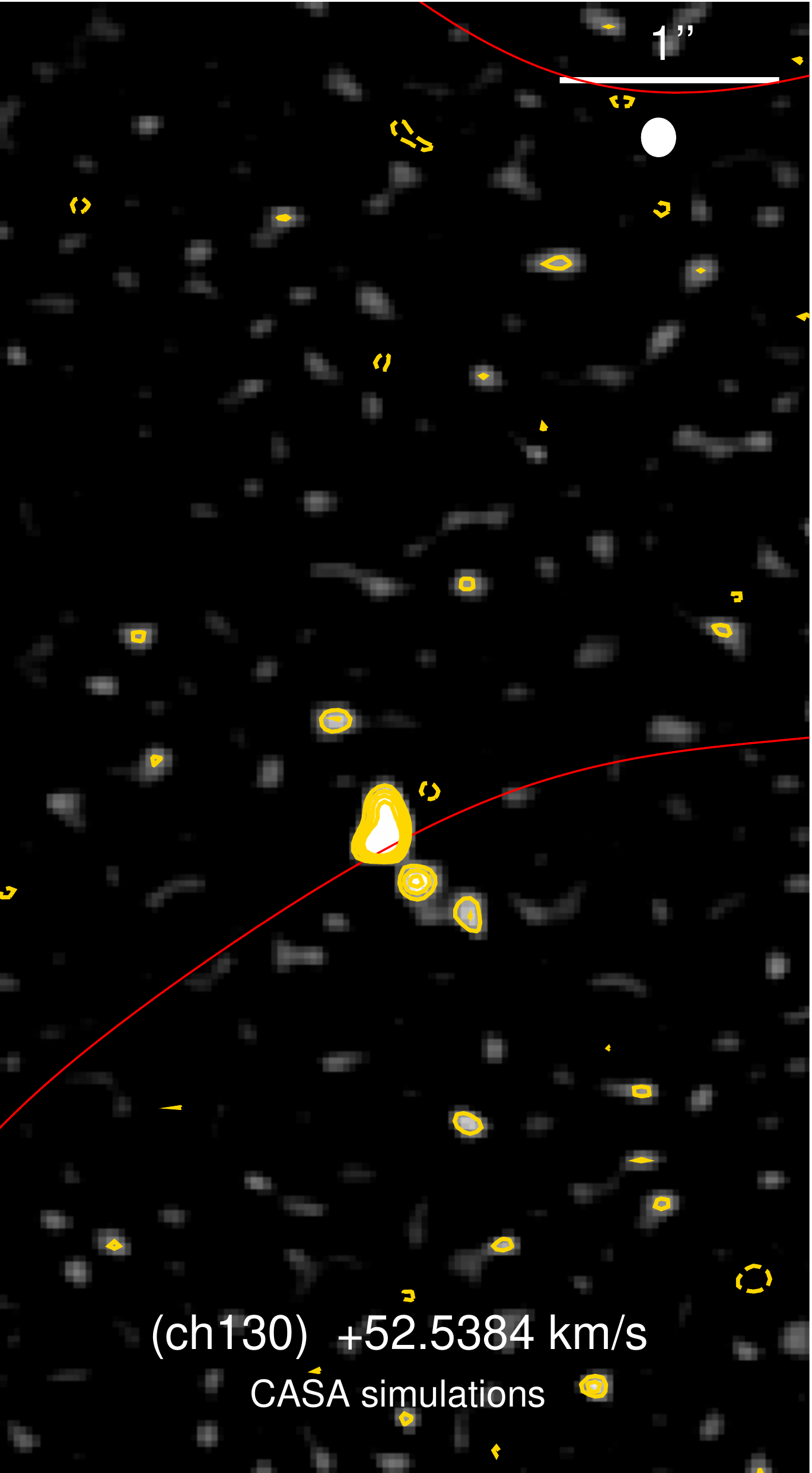}
\includegraphics[width=0.245\textwidth,clip]{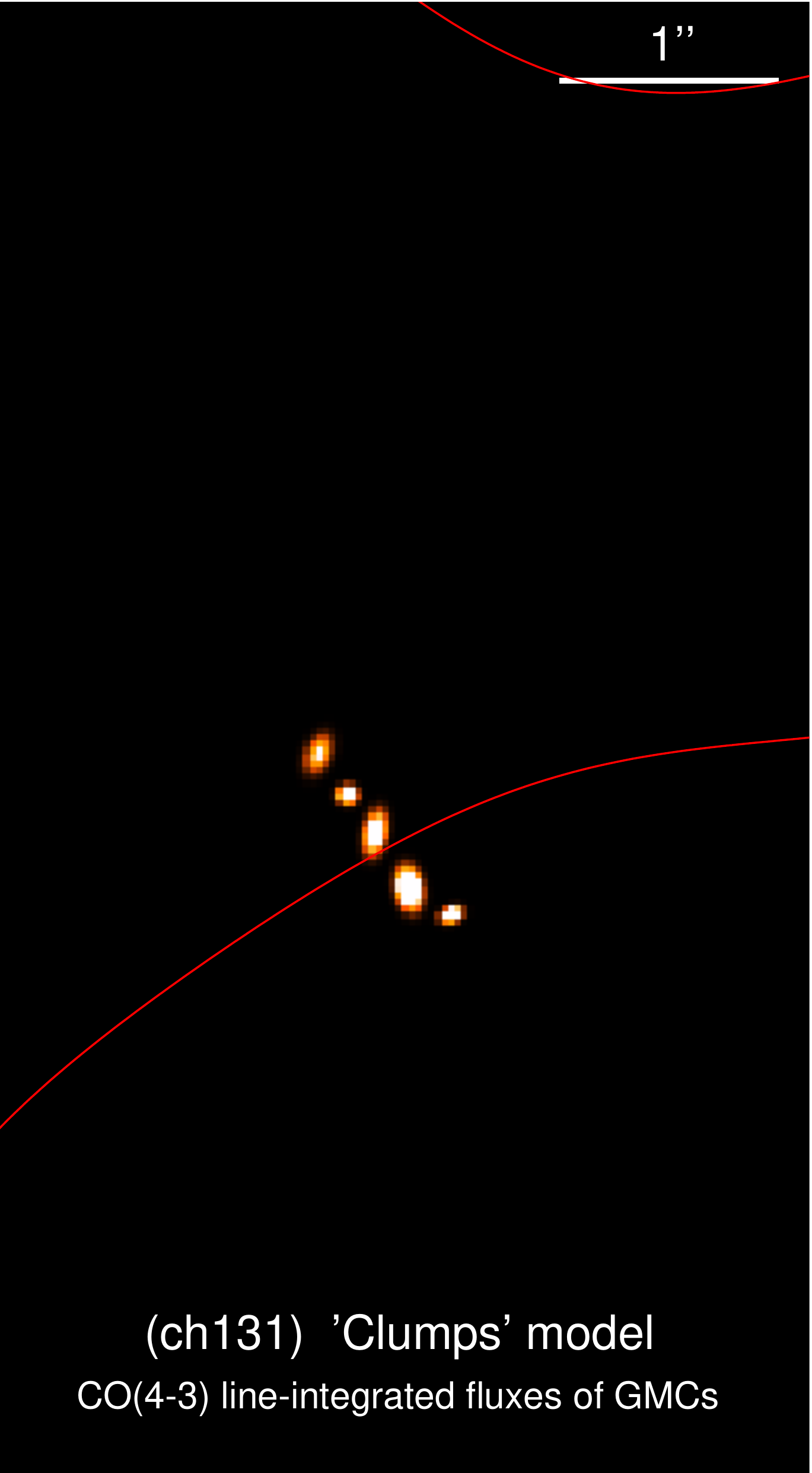}
\includegraphics[width=0.245\textwidth,clip]{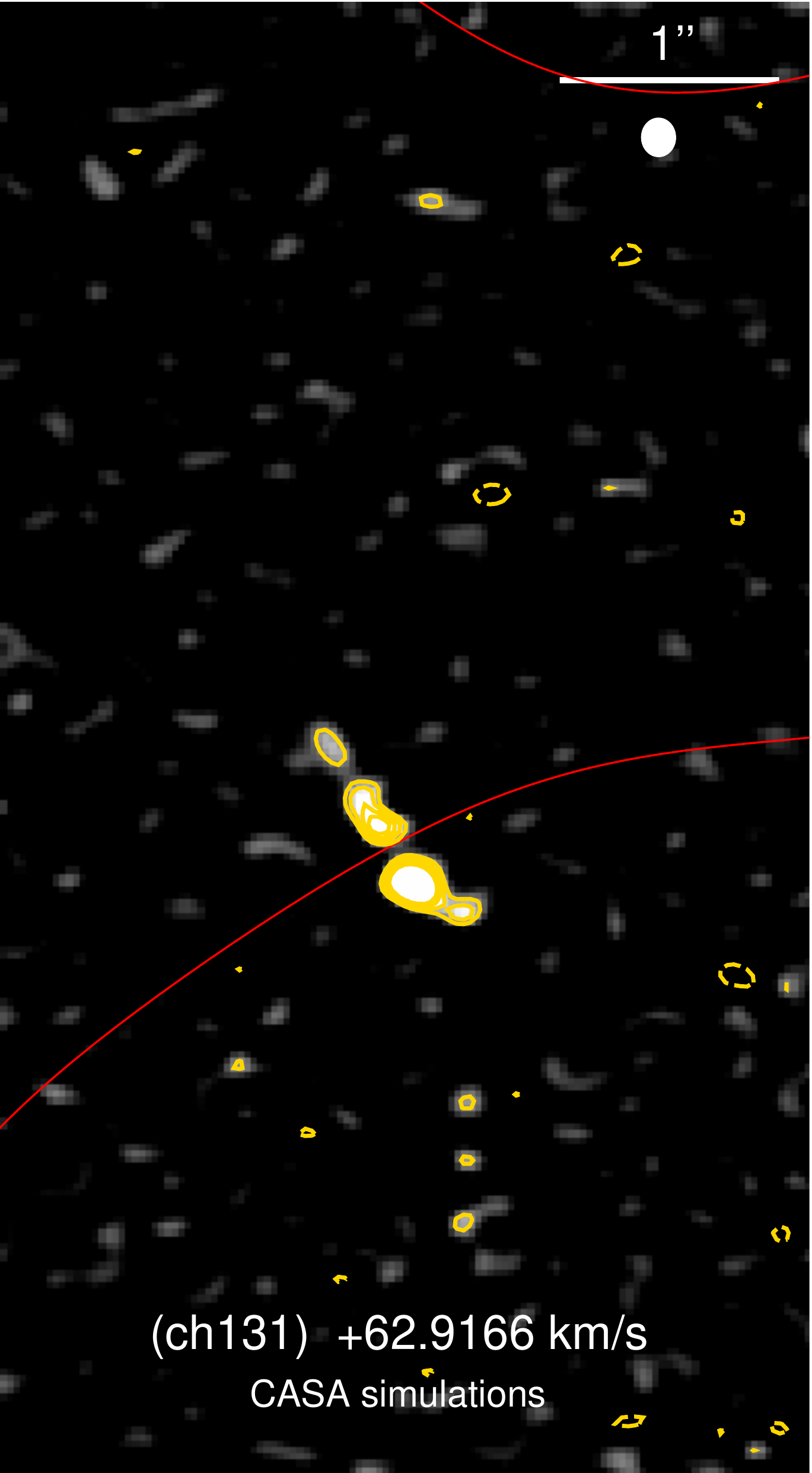}
\includegraphics[width=0.243\textwidth,clip]{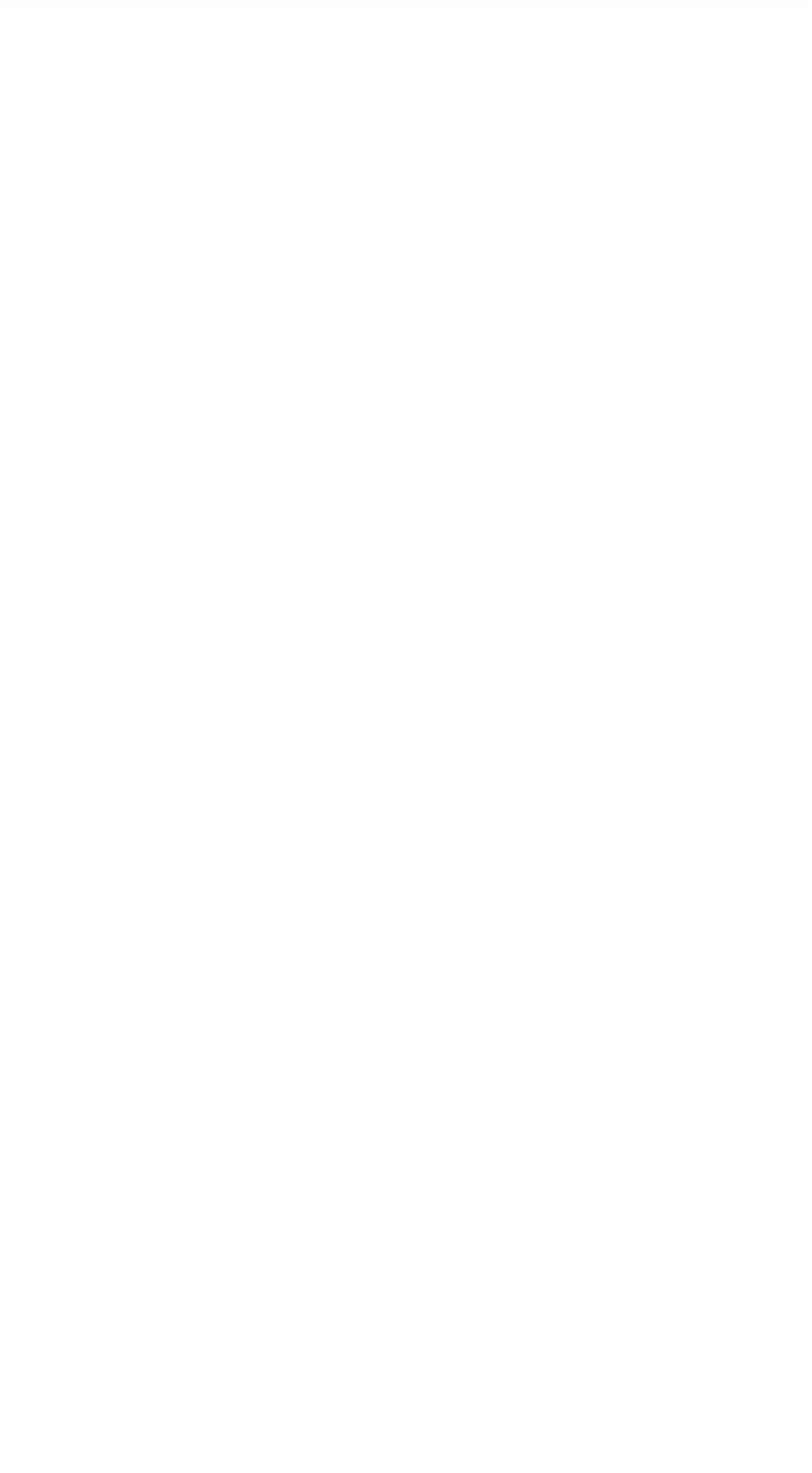}
\includegraphics[width=0.243\textwidth,clip]{figure-blank.pdf}
\caption{{\it ALMA} visibility simulations of CO(4--3) clumps placed in the channel intensity maps of $10.3782~\rm km~s^{-1}$ at the locations of the GMGs identified in the A521-sys1 galaxy and with their measured properties (fluxes and $a$, $b$ elliptical sizes). We show the corresponding sky model on the left, and the CASA simulated intensity maps with the RMS noise comparable to the real {\it ALMA} observations on the right for three consecutive channels encompassing the central region of the A521-sys1 galaxy. The modelled CO(4--3) clumps are detected at $>4-6\sigma$ above the RMS noise in the simulated channel maps, namely at similar significance levels as the GMCs detected in the real channel maps (see Fig.~\ref{fig:Appendix}). For different noise realisations, the clumps remain detected in the simulated channel maps at comparable significance levels. The overlaid gold contours start at $\pm 3\sigma$, in steps of $1\sigma$ ($\rm RMS = 0.003~Jy~beam^{-1}~km~s^{-1}$); dashed gold contours for negative values.
The critical line of our lens model at the redshift of the A521-sys1 galaxy, $z=1.043$, is shown by the red solid line. The $0.18''\times 0.15''$ synthesized beam of the simulated data at the position angle of $-89^{\circ}$ is represented by the white filled ellipse. We give a reference scale of 1 arcsec.}
\label{fig:Appendix-simu_clumps}
\end{figure*}

In the second sky model, we created a data cube with also a $10.3782~\rm km~s^{-1}$ sampling and we placed in the channel intensity maps CO(4--3) clumps at the locations of the GMGs identified in the A521-sys1 galaxy and with their measured properties (fluxes and $a$, $b$ elliptical sizes) in the image plane. The modelled CO(4--3) clumps are detected at $>4-6\sigma$ above the RMS noise in the simulated channel intensity maps, namely at similar significance levels as the GMCs detected in the real channel intensity maps as shown in Fig.~\ref{fig:Appendix-simu_clumps}. For different noise realisations in our CASA visibility simulations, the clumps remained detected in the simulated channel maps at comparable significance levels. 

These two sets of CASA visibility simulations demonstrate that clumps that benefit from a high CO(4--3) surface brightness can be detected in channel intensity maps. This is not the case if the measured total CO(4--3) line-integrated flux is distributed in a smooth exponential disc that globally suffers from a much lower surface brightness than clumps.


\bsp	
\label{lastpage}
\end{document}